\documentclass[11pt]{article}
\usepackage[totalwidth=460truept,totalheight=600truept]{geometry}
\usepackage{latexsym,amssymb,amsmath,graphicx,accents,eucal,slashed,subfigure}
\usepackage{dsfont}
\usepackage[T1]{fontenc}
\usepackage{hyperref}

\def\theequation{\arabic{section}.\arabic{equation}}

\renewcommand{\theequation}{\thesection.\arabic{equation}}
\global\arraycolsep=1truept

\newtheorem{theorem}{Theorem}
\numberwithin{equation}{section}
\renewcommand{\theequation}{\arabic{section}.\arabic{equation}}

\begin{document}

\bigskip \hfill IFUP-TH 2012/10

\vskip 1.4truecm

\begin{center}
{\huge \textbf{Master Functional And Proper Formalism}}

\vskip .5truecm

{\huge \textbf{For Quantum Gauge Field Theory}}

\vskip 1truecm

\textsl{Damiano Anselmi} \vskip .2truecm

\textit{Dipartimento di Fisica ``Enrico Fermi'', Universit\`{a} di Pisa, }

\textit{Largo B. Pontecorvo 3, I-56127 Pisa, Italy,}

\vskip .2truecm

damiano.anselmi@df.unipi.it

\vskip 1.5truecm

\textbf{Abstract}
\end{center}

\medskip

{\small We develop a general field-covariant approach to quantum gauge
theories. Extending the usual set of integrated fields and external sources
to ``proper'' fields and sources, which include partners of the composite
fields, we define the master functional }$\Omega ${\small , which collects
one-particle irreducible diagrams and upgrades the usual }$\Gamma ${\small %
-functional in several respects. The functional }$\Omega ${\small \ is
determined from its classical limit applying the usual diagrammatic rules to
the proper fields. Moreover, it behaves as a scalar under the most general
perturbative field redefinitions, which can be expressed as linear
transformations of the proper fields. We extend the Batalin-Vilkovisky
formalism and the master equation. The master functional satisfies the
extended master equation and behaves as a scalar under canonical
transformations. The most general perturbative field redefinitions and
changes of gauge-fixing can be encoded in proper canonical transformations,
which are linear and do not mix integrated fields and external sources.
Therefore, they can be applied as true changes of variables in the
functional integral, instead of mere replacements of integrands. This
property overcomes a major difficulty of the functional }$\Gamma ${\small .
Finally, the new approach allows us to prove the renormalizability of gauge
theories in a general field-covariant setting. We generalize known
cohomological theorems to the master functional and show that when there are
no gauge anomalies all divergences can be subtracted by means of parameter
redefinitions and proper canonical transformations. }

\vskip 1truecm

\vfill\eject

\section{Introduction}

\setcounter{equation}{0}

The renormalization of gauge theories is efficiently studied using the
Batalin-Vilkovisky formalism \cite{bata} and the dimensional-regularization
technique. First, sources $K_{A}$ are coupled to the gauge transformations $%
R^{A}(\Phi )$ of the fields $\Phi ^{A}$. Then the pairs $\Phi ^{A}$-$K_{A}$
are viewed as conjugate variables, and a notion of antiparentheses $(X,Y)$
for local and non-local functionals $X$ and $Y$ is introduced. The master
equations $(S,S)=0$ and $(\Gamma ,\Gamma )=0$ satisfied by the action $S$
and the generating functional $\Gamma $ of one-particle irreducible diagrams
keep track of gauge invariance through radiative corrections and
counterterms. Divergences proportional to the field equations or of
gauge-fixing type are removed by means of canonical transformations, while
divergences of all other types are removed by means of parameter
redefinitions.

The canonical transformations are the redefinitions of $\Phi $ and $K$ that
preserve the antiparentheses. In simple power-counting renormalizable
theories, such as ordinary Yang-Mills theory, the renormalization of fields
and sources is multiplicative. However, in more complicated, renormalizable
or non-renormalizable, theories, such as Yang-Mills theory with an unusual
action, or with composite fields turned on, as well as effective field
theories and gravity, the canonical transformations that subtract
divergences can be non-linear \cite{thooftveltman} and mix $\Phi $ and $K$
in non-trivial ways. It is legitimate to replace the integrated fields $\Phi 
$ with (perturbatively local) functions of $\Phi $ and $K$, because this
operation is just a change of variables in the functional integral.
Nevertheless, it is not legitimate to replace an external source $K$ with a
function that depends on the integrated fields $\Phi $. Therefore, the
formalism we are accustomed to is not completely satisfactory. It works only
if canonical transformations are meant as mere replacements of integrands,
not true changes of field variables in the functional integral. This means
that they are understood as changes of variables only at the level of the
action $S$. However, the term $\int \Phi ^{A}J_{A}$ that appears in the
exponent of the $Z$-integrand is not transformed, but just replaced by brute
force with the one of the new variables, $\int \Phi ^{\prime \hspace{0.01in}%
A}J_{A}^{\prime }$.

When the entire $J$-dependence is encoded in the term $\int \Phi ^{A}J_{A}$
we say that the functional integral is written in the \textit{conventional
form}. A non-linear change of variables destroys the conventional form.
Replacements bypass this fact restoring that form by brute force. Thus,
replacements allow us to ``jump'' from the generating functional associated
with the action $S$ to the one associated with the transformed action $%
S^{\prime }$, but cannot be straightforwardly used to write identities
relating the generating functionals before and after the transformation.

A general field-covariant approach to quantum field theory has been
developed in ref.s \cite{fieldcov,masterf}. Due to the intimate relation
between composite fields and changes of field variables, sources $L_{I}$ for
the composite fields $\mathcal{O}^{I}(\varphi )$ are introduced, besides the
usual sources $J$ for the elementary fields $\varphi $. Then, suitably
improved $Z$- and $W$-functionals are defined, so that the Legendre
transform of $W$ with respect to both $J$ and $L$ exists, and defines the
so-called master functional $\Omega (\Phi ,N)$. The master functional is a
generating functional of one-particle irreducible diagrams and behaves as a
scalar under general changes of field variables. Moreover, every field
redefinition is encoded in a linear transformation of $\Phi $ and $N$. In a
particularly convenient ``proper'' approach, where the integrated fields $%
\varphi $ are extended to a set of proper fields $\varphi $ and $N_{S}$,
such that $\Phi =\langle \varphi \rangle $ and $N=\langle N_{S}\rangle $,
the radiative corrections to $\Omega$ can be derived from its classical
limit following rules analogous to the ones we are accustomed to. Moreover,
the conventional form of the functional integral is manifestly preserved
during a general change of field variables, so replacements and true changes
of field variables are practically the same thing.

If we want a truly variable-independent approach to gauge field theory we
must generalize the ideas of ref. \cite{masterf} and define the master
functional $\Omega $ in a way that is compatible with gauge invariance. In
particular, $\Omega $ must satisfy it own master equation. In this paper we
show how this is done.

We first extend the notion of antiparentheses and the Batalin-Vilkovisky
formalism to the composite-field sector, and prove that $\Omega $ satisfies
a master equation in this extended sense. We also extend the proper
formulation of $\Omega $. In particular, there exist \textit{proper }%
canonical transformations, which are linear redefinitions of the proper
fields and the proper sources that encode the most general field
redefinitions and changes of gauge-fixing. The external sources transform
without mixing with the integrated fields. Thus, proper canonical
transformations can be safely implemented as changes of field variables in
the functional integral.

The proper formalism allows us to study the renormalizability of gauge
theories in a general setting. We assume that there are no gauge anomalies
and the gauge algebra closes off shell. We show that when known
cohomological theorems \cite{coho} apply they can be generalized to the
master functional with a relatively small effort. Then the classical and
bare actions can be extended till they include enough independent parameters
so that all divergences proportional to the field equations and of
gauge-fixing type are subtracted by means of proper canonical
transformations and all other divergences are subtracted by means of
parameter redefinitions. Clearly, the master equation is preserved at every
step of the subtraction.

For definiteness, we work using the Euclidean notation, but no results
depend on this choice.

The paper is organized as follows. In section 2 we define the master
functional for gauge theories and extend the antiparentheses and the master
equation to the composite-field sector. In section 3 we develop the proper
formalism, and in section 4 we study the proper canonical transformations.
In sections 5, 6 and 7 we study the renormalization of gauge theories in the
general field-covariant approach. In section 5, we prove some preliminary
results, subtracting divergences as they appear, without checking if they
can be reabsorbed inside canonical transformations and parameter
redefinitions. Then, in section 6, we generalize known cohomological
theorems to the master functional and prove that $\Omega $ can be
renormalized by means of parameter redefinitions and generic canonical
transformations. At this stage we do not know whether those canonical
transformations are proper or not, so we can only use them as replacements.
In section 7 we extend the classical and bare actions so that they still
satisfy their master equations and contain enough independent parameters to
subtract all divergences by means of parameter redefinitions and proper
canonical transformations. Section 8 contains the conclusions and in the
appendix we prove a cohomological theorem used in the paper.

\section{Master functional}

\setcounter{equation}{0}

In this section we generalize the master functional introduced in ref. \cite
{masterf} to gauge field theories, and extend the Batalin-Vilkovisky
formalism.

First we briefly recall the usual formalism. The set of fields $\{\Phi
^{A}\} $ includes the classical fields $\phi $, plus ghosts $C$, antighosts $%
\bar{C} $ and Lagrange multipliers $B$ for the gauge-fixings. An external
source $K_{A}$ is associated with each $\Phi ^{A}$, with statistics opposite
to the one of $\Phi ^{A}$. If $X$ and $Y$ are functionals of $\Phi $ and $K$
their antiparentheses are defined as 
\begin{equation}
(X,Y)\equiv \int \left( \frac{\delta _{r}X}{\delta \Phi ^{A}}\frac{\delta
_{l}Y}{\delta K_{A}}-\frac{\delta _{r}X}{\delta K_{A}}\frac{\delta _{l}Y}{%
\delta \Phi ^{A}}\right) ,  \label{usa}
\end{equation}
where the summation over $A$ is understood. The integral is made over
spacetime points associated with repeated indices. The master equation is $%
(S,S)=0$. Solving it with the ``boundary condition'' $S(\Phi ,K)=S_{c}(\phi
) $ at $C=\bar{C}=B=K=0$, where $S_{c}(\phi )$ is the classical action, we
get the solution $S(\Phi ,K)$ we start with to quantize the theory. We
assume that the gauge algebra closes off shell, so there exists a variable
frame where $S(\Phi ,K)$ is linear in $K$. Then we can define functions $%
\mathcal{S}(\Phi )$ and $R^{A}(\Phi )$, such that 
\begin{equation}
S(\Phi ,K)=\mathcal{S}(\Phi )-\int R^{A}(\Phi )K_{A}.  \label{buh}
\end{equation}
We assume that the source-independent action $\mathcal{S}(\Phi )$ is already
gauge-fixed. The generating functional $Z$ reads 
\[
Z(J,K)=\int [\mathrm{d}\Phi ]\exp \left( -S(\Phi ,K)+\int \Phi
^{A}J_{A}\right) =\exp W(J,K). 
\]
Finally, the generating functional $\Gamma (\Phi ,K)$ of one-particle
irreducible diagrams is the Legendre transform of $W(J,K)$ with respect to $%
J $, the sources $K$ being spectators.

For example, in pure non-Abelian Yang-Mills theory we have $\Phi
^{A}=(A_{\mu }^{a},C^{a},\bar{C}^{a},B^{a})$, $K_{A}=(K_{a}^{\mu
},K_{C}^{a},K_{\bar{C}}^{a},K_{B}^{a})$ and 
\begin{eqnarray*}
S(\Phi ,K) &=&\int \left( \frac{1}{4}F_{\mu \nu }^{a\ 2}-\frac{\lambda }{2}%
(B^{a})^{2}+B^{a}\partial \cdot A^{a}-\bar{C}^{a}\partial _{\mu }D_{\mu
}C^{a}\right) \\
&&-\int D_{\mu }C^{a}K_{\mu }^{a}+\frac{g}{2}\int
f^{abc}C^{b}C^{c}K_{C}^{a}-\int B^{a}K_{\bar{C}}^{a}{,}
\end{eqnarray*}
where $F_{\mu \nu }^{a}=\partial _{\mu }A_{\nu }^{a}-\partial _{\nu }A_{\mu
}^{a}+gf^{abc}A_{\mu }^{b}A_{\nu }^{c}$ is the field strength, $D_{\mu
}C^{a}=\partial _{\mu }C^{a}+gf^{abc}A_{\mu }^{b}C^{c}$ is the covariant
derivative of the ghosts and $f^{abc}$ are the structure constants of the
Lie algebra.

Starting from a variable frame where the structure (\ref{buh}) holds is not
a restrictive assumption, because the form (\ref{buh}) is preserved by all
canonical transformations whose generating functions $F(\Phi ,K^{\prime })$
are linear in the sources, 
\[
F(\Phi ,K^{\prime })=\int (\Phi ^{A}+U^{A}(\Phi ))K_{A}^{\prime }+\Psi (\Phi
), 
\]
which contain the most general $\Phi $-redefinitions and changes of
gauge-fixings. Note, however, that these canonical transformations turn the
sources into functions of the fields, so they cannot be used as changes of
variables in the functional integral, but must be treated as replacements of
integrands.

The master functional allows us to solve this and other problems, and
formulate quantum field theory in a general field-covariant way. A
non-linear change of field variables is intimately related to composite
fields. We need to consider all sorts of composite fields, not only
gauge-invariant ones, because a field redefinition is in no way restricted
by gauge invariance. However, gauge-non-invariant composite fields make the
action violate the master equation, so they must be introduced in a clever
way. We are going to include composite fields with arbitrary ghost number,
and both bosonic and fermionic statistics, made with the fields $\Phi $ and
their derivatives. Thus, the first thing to do it to add a term $\int 
\mathcal{O}^{I}(\Phi )L_{I}$ to the exponent of the $Z$-integrand, where $\{%
\mathcal{O}^{I}\}$ is a basis of composite fields and $L_{I}$ denotes the
external source coupled to $\mathcal{O}^{I}$.

Since this extension violates the master equation, we need to further modify
the exponent, so that it also incorporates the gauge transformations 
\[
R_{\mathcal{O}}^{I}(\Phi )\equiv \int R^{A}(\Phi )\frac{\delta _{l}\mathcal{O%
}^{I}}{\delta \Phi ^{A}} 
\]
of the composite fields $\mathcal{O}^{I}(\Phi )$. We couple such functions
with new sources $H_{I}$. We must pay attention to the statistics of
composite fields and sources. By convention, we always place the sources to
the right, so the new exponent of the $Z$-integrand becomes 
\begin{equation}
-\mathcal{S}(\Phi )+\int \left( \Phi ^{A}J_{A}+R^{A}(\Phi )K_{A}+\mathcal{O}%
^{I}(\Phi )L_{I}+R_{\mathcal{O}}^{I}(\Phi )H_{I}\right) ,  \label{stutto}
\end{equation}
up to further extensions that we describe later, where summations over $I$
are also understood. Note that the composite fields $R_{\mathcal{O}}^{I}$s
appear twice: once together with the source $H_{I}$, and once together with
some sources $L_{\bar{I}}$. This redundancy is harmless, actually useful for
some purposes explained below. Moreover, the sources $L_{I}$ and $H_{I}$
play different roles in the Legendre transform that defines $\Omega $. It is
useful to rewrite expression (\ref{stutto}) as 
\begin{equation}
-S_{H}(\Phi ,K,H)+\int \left( \Phi ^{A}J_{A}+\mathcal{O}^{I}L_{I}\right) ,
\label{newa}
\end{equation}
where we have defined a new extended action 
\begin{equation}
S_{H}(\Phi ,K,H)=\mathcal{S}(\Phi )-\int \left( R^{A}(\Phi )K_{A}+R_{%
\mathcal{O}}^{I}(\Phi )H_{I}\right)  \label{newext}
\end{equation}
that obviously satisfies the usual master equation $(S,S)=0$. The last two
terms of (\ref{newa}), instead, both violate the master equation, which is
why we treat them on the same footing.

Observe that equating the sources $H$ to constants we can parametrize the
most general gauge-fixing, since the last term of (\ref{newext}) is of the
form $(S_{H},\Psi )$ with $\Psi =-\int \mathcal{O}^{I}H_{I}$. Moreover, if
we set the sources $L_{I}$ of gauge-invariant composite fields equal to
constants we can add interactions, thereby modifying the classical action at
will. Indeed, the techniques we are developing allow us to treat the most
general theory, express it using arbitrary field variables and gauge-fix it
with the most general gauge-fixing.

In ref.s \cite{fieldcov,masterf} a linear approach was defined, where the
action is extended to include higher-powers of the sources, so that
divergences are subtracted by means of field redefinitions, parameter
redefinitions and linear source redefinitions. Here similar results are less
straightforward to achieve, and we must proceed gradually. For the moment a
sufficiently general extended solution of the master equation is 
\begin{equation}
S_{H}(\Phi ,K,L,H)=\mathcal{S}(\Phi )-\int \left( R^{A}(\Phi )K_{A}+R_{%
\mathcal{O}}^{I}(\Phi )H_{I}\right) -\int \tau _{vI}\mathcal{N}^{v}(L)%
\mathcal{O}_{\text{inv}}^{I}(\Phi ),  \label{gullo}
\end{equation}
where $\mathcal{O}_{\text{inv}}^{I}$ are the gauge-invariant composite
fields (namely they satisfy $(S,\mathcal{O}_{\text{inv}}^{I})=(S_{H},%
\mathcal{O}_{\text{inv}}^{I})=0$), the $\tau _{vI}$s are constants and $%
\mathcal{N}^{v}(L)=\mathcal{O}(L^{2})$ are a basis of independent local
monomials, at least quadratic in $L_{I}$, that can be constructed with the
sources $L$ and their derivatives. Clearly, (\ref{gullo}) satisfies the
master equation $(S_{H},S_{H})=0$.

At this stage we cannot guarantee that all divergences can be subtracted
redefining the fields, sources and parameters appearing in (\ref{gullo}),
and that the sources redefinitions are linear. These problems are treated in
sections 6 and \ref{RGren}, where further extensions are developed. The
important fact is that (\ref{gullo}) is general enough to make room for the
improvement term met in ref. \cite{masterf}, which for later convenience we
write here as 
\begin{equation}
T(L)=\frac{1}{2}\int L_{I}(\tilde{A}^{-1})^{IJ}L_{J},  \label{improfin}
\end{equation}
where $A_{IJ}$ are constants ($c$-numbers) and $\tilde{A}$ is the $A$%
-transpose. These constants must be restricted according to the statistics
of the sources $L$, their ghost numbers and the other global symmetries
satisfied by the theory. The term (\ref{improfin}) is just a contribution to 
$\int \tau _{v0}\mathcal{N}^{v}$, which multiplies the identity operator,
and allows us to define the Legendre transform of $W$ with respect to $L$.
As usual, we shift $\int \tau _{v0}\mathcal{N}^{v}(L)$ by $T(L)$ and assume
that the new $\int \tau _{vI}\mathcal{N}^{v}\mathcal{O}_{\text{inv}}^{I}$ is
perturbative with respect to $T(L)$. The action becomes 
\begin{equation}
S_{H}(\Phi ,K,L,H)=\mathcal{S}(\Phi )-\int \left( R^{A}(\Phi )K_{A}+R_{%
\mathcal{O}}^{I}(\Phi )H_{I}\right) -\frac{1}{2}\int L_{I}(\tilde{A}%
^{-1})^{IJ}L_{J}-\int \tau _{vI}\mathcal{N}^{v}(L)\mathcal{O}_{\text{inv}%
}^{I}(\Phi ).  \label{gullo2}
\end{equation}
Again, $(S_{H},S_{H})=0$.

The generating functionals $Z$ and $W$ are defined as 
\begin{equation}
Z(J,K,L,H)=\mathrm{e}^{W(J,K,L,H)}=\int [\mathrm{d}\Phi ]\ \mathrm{\exp }%
\left( -S_{H}(\Phi ,K,L,H)+\int \left( \Phi ^{A}J_{A}+\mathcal{O}%
^{I}L_{I}\right) \right)  \label{zjkhlv}
\end{equation}
and are invariant under the transformation 
\[
\tau J_{A}=0,\qquad \tau K_{A}=(-1)^{\varepsilon _{A}}J_{A},\qquad \tau
L_{I}=0,\qquad \tau H_{I}=(-1)^{\varepsilon _{I}}L_{I}, 
\]
where $\varepsilon _{A}$ and $\varepsilon _{I}$ are the statistics of $\Phi
^{A}$ and $\mathcal{O}^{I}$, respectively. The proof of this invariance is
as follows. Apply the operator $\delta _{\tau }=\xi \tau $ to the $Z$%
-functional (\ref{zjkhlv}), where $\xi $ is a constant anticommuting
parameter. Using (\ref{gullo2}) we see that 
\begin{equation}
\delta _{\tau }W=-\xi \int \langle R^{A}J_{A}+R_{\mathcal{O}%
}^{I}L_{I}\rangle .  \label{bugi}
\end{equation}
These contributions can be canceled making the change of variables 
\begin{equation}
\Phi ^{A}\rightarrow \Phi ^{A}+\xi R^{A}=\Phi ^{A}+\xi (S_{H},\Phi ^{A})
\label{chv2}
\end{equation}
in (\ref{zjkhlv}). Indeed, (\ref{chv2}) affects only $\int \left( \Phi
^{A}J_{A}+\mathcal{O}^{I}L_{I}\right) $ by an amount opposite to (\ref{bugi}%
). Thus, 
\begin{equation}
\tau Z(J,K,L,H)=0,\qquad \tau W(J,K,L,H)=0.  \label{usta}
\end{equation}

We define the master functional $\Omega $ as the Legendre transform of $W$
with respect to $J$ and $L$, where $K$ and $H$ remain inert. We have 
\begin{equation}
\Omega (\Phi ,K,N,H)=-W(J,K,L,H)+\int \left( \Phi
^{A}J_{A}+N^{I}L_{I}\right) ,  \label{hug}
\end{equation}
where 
\begin{eqnarray}
\Phi ^{A} &=&\frac{\delta _{r}W}{\delta J_{A}},\qquad N^{I}=\frac{\delta
_{r}W}{\delta L_{I}},\qquad \frac{\delta _{r}W}{\delta K_{A}}=-\frac{\delta
_{r}\Omega }{\delta K_{A}},  \nonumber \\
J_{A} &=&\frac{\delta _{l}\Omega }{\delta \Phi ^{A}},\qquad L_{I}=\frac{%
\delta _{l}\Omega }{\delta N^{I}},\qquad \frac{\delta _{r}W}{\delta H_{I}}=-%
\frac{\delta _{r}\Omega }{\delta H_{I}}.  \label{assur}
\end{eqnarray}

Setting $K_{A}=H_{I}=L_{I}=0$ we switch off the composite-field sector and
the sector of gauge transformations. Instead, setting $H_{I}=L_{I}=0$ the
master functional $\Omega $ reduces to the $\Gamma $-functional $\Gamma
(\Phi ,K)$. In particular, the solutions of the conditions $L_{I}=\delta
_{l}\Omega /\delta N^{I}=0$ express $N^{I}$ as functions of the fields $\Phi 
$ and the sources $K$, $H$. Using $N^{I}=\delta _{r}W/\delta L_{I}$ and
formula (\ref{zjkhlv}) at $L=0$ we have $N^{I}=\langle \mathcal{O}%
^{I}\rangle _{L=0}$, therefore 
\[
\Gamma (\Phi ,K)=\Omega (\Phi ,K,\langle \mathcal{O}^{I}\rangle _{L=H=0},0). 
\]

\medskip

Now we extend the Batalin-Vilkovisky formalism. Given two functionals $X$
and $Y$ of $\Phi $, $K$, $N$ and $H$, we define their \textit{squared
antiparentheses} $\left\lfloor X,Y\right\rfloor $ as 
\[
\left\lfloor X,Y\right\rfloor \equiv \int \left( \frac{\delta _{r}X}{\delta
\Phi ^{A}}\frac{\delta _{l}Y}{\delta K_{A}}+\frac{\delta _{r}X}{\delta N^{I}}%
\frac{\delta _{l}Y}{\delta H_{I}}-\frac{\delta _{r}X}{\delta K_{A}}\frac{%
\delta _{l}Y}{\delta \Phi ^{A}}-\frac{\delta _{r}X}{\delta H_{I}}\frac{%
\delta _{l}Y}{\delta N^{I}}\right) . 
\]
In practice, the squared antiparentheses are made of the usual
antiparentheses plus similar ones for the pair of conjugate variables $N^{I}$%
-$H_{I}$. Because of this, the squared antiparentheses satisfy the usual
properties 
\begin{eqnarray}
&&\qquad \lfloor Y,X\rfloor =-(-1)^{(\varepsilon _{X}+1)(\varepsilon
_{Y}+1)}\lfloor X,Y\rfloor ,  \nonumber \\
&&(-1)^{(\varepsilon _{X}+1)(\varepsilon _{Z}+1)}\lfloor X,\lfloor
Y,Z\rfloor \rfloor +\text{cyclic permutations}=0,  \label{Jid}
\end{eqnarray}
and $\varepsilon _{\lfloor X,Y\rfloor }=\varepsilon _{X}+\varepsilon _{Y}+1$%
, where $\varepsilon _{X}$ denotes the statistics of the functional $X$.
Moreover, 
\begin{equation}
\lfloor F,F\rfloor =0,\qquad \lfloor B,B\rfloor =-2\int \left( \frac{\delta
_{r}B}{\delta K_{A}}\frac{\delta _{l}B}{\delta \Phi ^{A}}+\frac{\delta _{r}B%
}{\delta H_{I}}\frac{\delta _{l}B}{\delta N^{I}}\right) ,  \label{bubu}
\end{equation}
if $F$ is a fermionic functional and $B$ is a bosonic one. Finally, the
Jacobi identity (\ref{Jid}) ensures that 
\begin{equation}
\lfloor X,\lfloor X,X\rfloor \rfloor =0  \label{wezum}
\end{equation}
for every functional $X$.

Now, using (\ref{assur}) and the second of (\ref{bubu}) we can rewrite 
\[
\tau W=-\int \frac{\delta _{r}W}{\delta K_{A}}J_{A}-\int \frac{\delta _{r}W}{%
\delta H_{I}}L_{I}=0 
\]
as 
\begin{equation}
\tau W=\int \frac{\delta _{r}\Omega }{\delta K_{A}}\frac{\delta _{l}\Omega }{%
\delta \Phi ^{A}}+\int \frac{\delta _{r}\Omega }{\delta H_{I}}\frac{\delta
_{l}\Omega }{\delta N^{I}}=-\frac{1}{2}\lfloor \Omega ,\Omega \rfloor =0.
\label{milpot}
\end{equation}
We have thus proved that the master functional $\Omega $ satisfies the
master equation $\lfloor \Omega ,\Omega \rfloor =0$.

Canonical transformations can be extended defining them as the redefinitions
of $\Phi $, $K$, $N$ and $H$ that preserve the squared antiparentheses. They
can be encoded in generating functions $F(\Phi ,K^{\prime },N,H^{\prime })$
such that 
\[
\Phi ^{A\hspace{0.01in}\prime }=\frac{\delta F}{\delta K_{A}^{\prime }}%
,\qquad K_{A}=\frac{\delta F}{\delta \Phi ^{A}},\qquad N^{I\hspace{0.01in}%
\prime }=\frac{\delta F}{\delta H_{I}^{\prime }},\qquad H_{I}=\frac{\delta F%
}{\delta N^{I}}. 
\]

\section{Proper formulation}

\setcounter{equation}{0}

In this section we extend the proper formalism of ref. \cite{masterf} to
gauge theories, which allows us to derive several properties in more
economic and more general ways. For example, several results derived so far
rely on the particular form (\ref{gullo2}) of the action $S_{H}$. Now we can
drop that assumption. We still assume that the $L$-dependence of $S_{H}$ is
made of $-T(L)$ plus corrections that must be treated perturbatively with
respect to the improvement term. Moreover, the generating functionals are
still defined by (\ref{zjkhlv}) and (\ref{hug}). Next, define the $S_{L}$%
-action 
\begin{equation}
S_{L}(\Phi ,K,N,H,L)=S_{H}(\Phi ,K,L,H)+\int \tilde{N}^{I}L_{I},
\label{sln0}
\end{equation}
where $\tilde{N}^{I}=N^{I}-\mathcal{O}^{I}(\Phi )$, and assume that it
satisfies the master equation 
\begin{equation}
\lfloor S_{L},S_{L}\rfloor =0.  \label{slnprop}
\end{equation}
In this equation the sources $L$ are mere spectators. It is easy to show
that if $S_{H}$ is given by (\ref{gullo2}) the action $S_{L}$ does satisfies
this identity.

Define the \textit{proper action} $S_{N}(\Phi ,K,N,H)$ by means of the
formula 
\begin{equation}
\exp \left( -S_{N}(\Phi ,K,N,H)\right) \equiv \int [\mathrm{d}L]\exp \left(
-S_{L}(\Phi ,K,N,H,L)\right) .  \label{expsn}
\end{equation}
The generating functionals can be expressed using the proper action. For
example, we have 
\begin{equation}
Z(J,K,L,H)=\int [\mathrm{d}\Phi \hspace{0.01in}\mathrm{d}N]\exp \left(
-S_{N}(\Phi ,K,N,H)+\int \Phi ^{A}J_{A}+\int N^{I}L_{I}\right) =\exp
W(J,K,L,H),  \label{ibll}
\end{equation}
because, using (\ref{sln0}) and (\ref{expsn}), this formula immediately
returns (\ref{zjkhlv}). Then the master functional can be calculated using (%
\ref{hug})-(\ref{assur}).

Observe that in this section $\Phi $ and $N$ denote the integrated proper
fields, arguments of the proper action $S_{N}$, while in the previous
section the same notation $\Phi $ and $N$ was used for the arguments of the
master functional $\Omega $. When we want to distinguish the ones from the
others, we call $\Phi _{\Omega }$, $N_{\Omega }$ the arguments of $\Omega $
and $\Phi _{S}$, $N_{S}$ the arguments of $S_{N}$. Thus, strictly speaking
we have $\Phi _{\Omega }=\langle \Phi _{S}\rangle $ and $N_{\Omega }=\langle
N_{S}\rangle $.

In the functional integral (\ref{expsn}) the $L$-propagators are those
determined by the improvement term (\ref{improfin}), therefore they are
proportional to the identity, and every other term of (\ref{expsn}) must be
treated perturbatively. It is easy to see, using the dimensional
regularization, that the integral of (\ref{expsn}) receives contributions
only from tree diagrams and can be calculated exactly using the saddle-point
approximation. Let $L_{I}=L_{I}^{*}(\Phi ,K,N,H)$ denote the solution of $%
\delta _{r}S_{L}/\delta L=0$, or 
\begin{equation}
\tilde{N}^{I}=-\frac{\delta _{r}}{\delta L_{I}}S_{H}(\Phi ,K,L,H).
\label{arrepo}
\end{equation}
Then 
\[
S_{N}(\Phi ,K,N,H)=S_{H}(\Phi ,K,L^{*},H)+\int \tilde{N}^{I}L_{I}^{*}. 
\]

In practice, $S_{N}$ coincides with the Legendre transform of $-S_{H}+\int 
\mathcal{O}^{I}L_{I}$ with respect to $L$. Because of this, we have the
relation 
\begin{equation}
\frac{\delta _{l}S_{N}}{\delta N^{I}}=L_{I}^{*}(\Phi ,K,N,H),  \label{bigo}
\end{equation}
so (\ref{arrepo}) is equivalent to 
\begin{equation}
L_{I}=\frac{\delta _{l}S_{N}}{\delta N^{I}}.  \label{fe}
\end{equation}

If $S_{L}$ satisfies (\ref{slnprop}) the proper action satisfies the \textit{%
proper master equation} 
\begin{equation}
\lfloor S_{N},S_{N}\rfloor =0.  \label{snprop}
\end{equation}
Indeed, observe that $S_{N}$ is just $S_{L}$ with $L$ equal to the solution $%
L^{*}$ of $\delta _{r}S_{L}/\delta L=0$. Every time we differentiate $S_{N}$
with respect to $\vartheta =\Phi ,K,N$ or $H$, we must pay attention to the
fact that now $L$ is no longer fixed, but depends on $\vartheta $. However,
the operation of setting $L=L^{*}$ and the differentiation with respect to $%
\vartheta $ can be freely interchanged on $S_{L}$, because 
\[
\frac{\delta _{r}S_{N}}{\delta \vartheta }=\frac{\delta _{r}}{\delta
\vartheta }\left( \left. S_{L}\right| _{L=L^{*}}\right) =\left. \frac{\delta
_{r}S_{L}}{\delta \vartheta }\right| _{L=L^{*}}+\left. \frac{\delta _{r}S_{L}%
}{\delta L_{I}}\right| _{L=L^{*}}\frac{\delta _{r}L_{I}^{*}}{\delta
\vartheta }=\left. \frac{\delta _{r}S_{L}}{\delta \vartheta }\right|
_{L=L^{*}}. 
\]
Here the integrations over spacetime points associated with repeated indices
are understood. Thus, (\ref{slnprop}) implies (\ref{snprop}). Moreover,
differentiating $\lfloor S_{L},S_{L}\rfloor =0$ with respect to $L_{I}$ we
have $\lfloor S_{L},\delta _{r}S_{L}/\delta L_{I}\rfloor =0$. Setting $%
L=L^{*}$ in this equation and using 
\[
0=\frac{\delta _{l}}{\delta \vartheta }\left( \left. \frac{\delta _{r}S_{L}}{%
\delta L_{I}}\right| _{L=L^{*}}\right) =\left. \frac{\delta _{l}\delta
_{r}S_{L}}{\delta \vartheta \delta L_{I}}\right| _{L=L^{*}}+\frac{\delta
_{l}L_{J}^{*}}{\delta \vartheta }\left. \frac{\delta _{l}\delta _{r}S_{L}}{%
\delta L_{J}\delta L_{I}}\right| _{L=L^{*}}, 
\]
we get 
\[
\lfloor S_{N},L_{J}^{*}\rfloor \left. \frac{\delta _{l}\delta _{r}S_{L}}{%
\delta L_{J}\delta L_{I}}\right| _{L=L^{*}}=0. 
\]
Now, $\delta _{l}\delta _{r}S_{L}/\delta L_{J}\delta L_{I}$ is a
perturbatively invertible matrix, because the improvement term is dominant
with respect to all other terms belonging to the $L$-sector. Therefore, we
can drop that matrix and conclude 
\begin{equation}
\lfloor S_{N},L_{I}^{*}\rfloor =0.  \label{cucu}
\end{equation}
This result was expected, because $L$ is gauge-invariant ($\lfloor
S_{L},L_{I}\rfloor =0$) and $L^{*}$ is determined by the gauge-invariant
equation $\delta _{r}S_{L}/\delta L=0$. Then the solution $L=L^{*}(\Phi
,K,N,H)$ must be gauge-invariant.

Before proceeding, we apply the construction just outlined to the case where 
$S_{H}$ is given by (\ref{gullo2}). Using (\ref{arrepo}) we find 
\begin{equation}
L_{I}^{*}=A_{IJ}\tilde{N}^{J}+\mathcal{O}(\tilde{N})\text{-perturbative
corrections},  \label{arrarrarra}
\end{equation}
thus the proper action has the form 
\begin{equation}
S_{N}(\Phi ,K,N,H)=\mathcal{S}(\Phi )-\int \left( R^{A}(\Phi )K_{A}+R_{%
\mathcal{O}}^{I}(\Phi )H_{I}\right) +\frac{1}{2}\int \tilde{N}^{I}A_{IJ}%
\tilde{N}^{J}+\int \rho _{vI}\mathcal{N}^{v}(\tilde{N})\mathcal{O}_{\text{inv%
}}^{I}(\Phi ),  \label{gullo3}
\end{equation}
for suitable parameters $\rho _{vI}$ equal to $\tau _{vI}$ plus perturbative
corrections. Observe that the combinations $\tilde{N}$ are gauge invariant,
namely $\lfloor S_{N},\tilde{N}^{I}\rfloor =0$.

It is also instructive to prove (\ref{snprop}) more directly in this case.
We do it using (\ref{expsn}). Observe that 
\[
\frac{\delta _{r}S_{N}}{\delta K_{A}}=-R^{A}(\Phi ),\qquad \frac{\delta
_{l}S_{N}}{\delta \Phi ^{A}}=\left\langle \frac{\delta _{l}S_{H}}{\delta
\Phi ^{A}}\right\rangle _{L}-\frac{\delta _{l}\mathcal{O}^{I}}{\delta \Phi
^{A}}\langle L_{I}\rangle _{L},\qquad \frac{\delta _{l}S_{N}}{\delta N^{I}}%
=\langle L_{I}\rangle _{L},\qquad \frac{\delta _{r}S_{N}}{\delta H_{I}}=-R_{%
\mathcal{O}}^{I}, 
\]
where $\langle \cdots \rangle _{L}$ denotes the expectation value in the
sense of the $L$-integral of (\ref{expsn}). Using (\ref{bubu}) we have 
\[
\lfloor S_{N},S_{N}\rfloor =2\int \left\langle R^{A}\frac{\delta _{l}S_{H}}{%
\delta \Phi ^{A}}\right\rangle _{L}=\langle (S_{H},S_{H})\rangle _{L}=0. 
\]

The perturbative expansion can be organized assuming \cite{masterf} 
\begin{eqnarray}
\lambda _{n_{l}} &=&\mathcal{O}(\delta ^{n_{l}-2}),\qquad L_{I}=\mathcal{O}%
(\delta ^{n_{I}-2}),\qquad A_{IJ}=\mathcal{O}(\delta
^{n_{I}+n_{J}-2}),\qquad \rho _{vI}=\mathcal{O}(\delta ^{n_{I}-n_{v}-1}), 
\nonumber \\
K^{A} &=&\mathcal{O}(\delta ^{n_{A}-2}),\qquad H_{I}=\mathcal{O}(\delta
^{n_{I}^{\prime }-2}),\qquad \tilde{N}_{S}^{I}=\mathcal{O}(\delta ^{-n_{I}}),
\label{assigne1}
\end{eqnarray}
where $\delta $ is some reference parameter $\ll 1$, $\lambda _{n_{l}}$ is
the coupling, or product of couplings, multiplying a monomial with $n_{l}$ $%
\Phi $-legs in $\mathcal{S}(\Phi )$, $n_{I}$ is such that $\mathcal{O}%
^{I}(\Phi \delta ^{-1},\lambda _{l}\delta ^{n_{l}-2})=\delta ^{-n_{I}}%
\mathcal{O}^{I}(\Phi ,\lambda )$, $n_{v}$ is the $\delta $-degree of $%
\mathcal{N}^{v}(L)$, while $n_{A}$ and $n_{I}^{\prime }$ are such $%
R^{A}(\Phi \delta ^{-1},\lambda _{l}\delta ^{n_{l}-2})=\delta
^{-n_{A}}R^{A}(\Phi ,\lambda _{l})$ and $R_{\mathcal{O}}^{I}(\Phi \delta
^{-1},\lambda _{l}\delta ^{n_{l}-2})=\delta ^{-n_{I}^{\prime }}R_{\mathcal{O}%
}^{I}(\Phi ,\lambda _{l})$. If we rescale every object by a factor $\delta
^{n}$, where $n$ it its $\delta $-degree, and in addition rescale $\Phi $ by 
$1/\delta $, then the proper action rescales as 
\begin{equation}
S_{N}(\Phi ,K,N,H)\rightarrow \frac{1}{\delta ^{2}}\bar{S}_{N}(\Phi ,K,N,H),
\label{jubi}
\end{equation}
where $\bar{S}_{N}$ has a factor $\delta $ for each $\rho $ and a factor $%
\delta ^{2}$ for each loop, but is $\delta $-independent everywhere else.
Formula (\ref{jubi}) ensures that radiative corrections are perturbative in $%
\delta $.

Summarizing, assumption (\ref{gullo2}) for $S_{H}$ is equivalent to
assumption (\ref{gullo3}) for $S_{N}$. Now we drop such assumptions again
and go back to the general case. We can take any classical proper action $%
S_{cN}$ that satisfies the following requirements:

\noindent 1) the perturbative expansion is defined around the usual $\Phi $%
-kinetic terms and the $N$-quadratic terms 
\begin{equation}
\frac{1}{2}\int N^{I}A_{IJ}N^{J},  \label{impric}
\end{equation}
2) differentiating $S_{cN}$ with respect to the sources $N^{I}$ and setting $%
\delta _{l}S_{cN}/\delta N^{I}$ to zero, the solutions $N^{I}(\Phi )$ at $%
K=H=0$ are a basis of local composite fields $\mathcal{O}_{c}^{I}(\Phi )$.

As usual, the renormalized action is denoted with $S_{N}$. The solutions $%
\tilde{N}^{I}\equiv N^{I}-\mathcal{O}_{\mathrm{R}}^{I}(\Phi )=0$ of $\delta
_{l}S_{N}/\delta N^{I}=0$ at $K=H=0$ define the renormalized composite
fields $\mathcal{O}_{\mathrm{R}}^{I}(\Phi )$. The composite-field sector and
the sector of gauge transformations can be switched off setting $\tilde{N}%
^{I}=K=H=0$. It is often convenient to write $S_{N}$ as a functional of $%
\Phi $, $K$, $\tilde{N}$ and $H$.

For the moment, the action $S_{N}$ may or may not satisfy the master
equation $\lfloor S_{N},S_{N}\rfloor =0$. Indeed, as we will see later in
some derivations it is useful to admit temporary violations of the master
equation.

\begin{theorem}
If the proper action $S_{N}$ satisfies the master equation $\lfloor
S_{N},S_{N}\rfloor =0$, then the master functional $\Omega $ satisfies the
master equation $\lfloor \Omega ,\Omega \rfloor =0$.
\end{theorem}

\textit{Proof}. We just need to recall that once we enlarge the set of
integrated fields from $\Phi $ to the proper fields $\Phi ,N$ and the set of
conjugate sources from $K$ to the proper sources $K,H$, then the master
functional $\Omega $ is the $\Gamma $-functional of the proper variables,
the squared antiparentheses are the usual antiparentheses, and the proper
master equation is the usual master equation. Thus the theorem follows from
the analogous statement that holds for usual $\Gamma $-functionals.

The theorem just proved can also be seen as a corollary of the following
more general theorem, concerning proper actions $S_{N}$ that do not
necessarily satisfy the master equation.

\begin{theorem}
The master functional $\Omega $ satisfies the identity 
\begin{equation}
\lfloor \Omega ,\Omega \rfloor =\langle \lfloor S_{N},S_{N}\rfloor \rangle .
\label{aididin}
\end{equation}
\end{theorem}

\textit{Proof}. Make the change of variables 
\[
\Phi \rightarrow \Phi +\xi \lfloor S_{N},\Phi \rfloor ,\qquad N\rightarrow
N+\xi \lfloor S_{N},N\rfloor , 
\]
in the functional integral (\ref{ibll}) that defines $Z$. Setting the
variation of $Z$ to zero we find the identity 
\[
0=-\frac{\xi }{2}\langle \lfloor S_{N},S_{N}\rfloor \rangle -\xi
\left\langle \int \frac{\delta _{r}S_{N}}{\delta K_{A}}J_{A}+\int \frac{%
\delta _{r}S_{N}}{\delta H_{I}}L_{I}\right\rangle . 
\]
Taking $J_{A}$ and $L_{I}$ out of the average and using 
\[
\left\langle \frac{\delta _{r}S_{N}}{\delta K_{A}}\right\rangle =-\frac{%
\delta _{r}W}{\delta K_{A}}=\frac{\delta _{r}\Omega }{\delta K_{A}}, 
\]
and similar identities for the $H$-derivatives, we obtain (\ref{aididin}).

\medskip

We have given a set of operations to construct $S_{N}$ from $S_{L}$. To
conclude this section we show that, despite the fact that $S_{L}$ depends on
more variables than $S_{N}$, we can invert those operations and go back to $%
S_{L}$. Given $S_{N}(\Phi ,K,N,H)$, solution of the master equation $\lfloor
S_{N},S_{N}\rfloor =0$, define $\tilde{S}_{L}$ as (minus) the Legendre
transform of $S_{N}$ with respect to $N$, namely 
\[
\tilde{S}_{L}(\Phi ,K,H,L)=S_{N}(\Phi ,K,N^{*},H)-\int N^{*I}L_{I}, 
\]
where $N^{I}=N^{*I}(\Phi ,K,H,L)$ are the solutions of (\ref{fe}). Then
define 
\begin{equation}
S_{L}(\Phi ,K,N,H,L)=\tilde{S}_{L}(\Phi ,K,H,L)+\int N^{I}L_{I}=\left.
S_{N}\right| _{N=N^{*}}+\int (N^{I}-N^{*I})L_{I}.  \label{slll}
\end{equation}
Inverting (\ref{sln0}) we can also work out $S_{H}(\Phi ,K,L,H)$, but we do
not need it right now. Clearly, the solution $L=L^{*}$ of $\delta
_{r}S_{L}/\delta L=0$ coincides with the solution $N=N^{*}$ of (\ref{fe}),
and we have $\left. S_{L}\right| _{L=L^{*}}=S_{N}$. It is easy to prove that 
$S_{L}$ satisfies $\lfloor S_{L},S_{L}\rfloor =0$, because for $\vartheta
=\Phi ,K$ or $H$ we have 
\begin{equation}
\frac{\delta _{l}S_{L}}{\delta \vartheta }=\left. \frac{\delta _{l}S_{N}}{%
\delta \vartheta }\right| _{N=N^{*}}+\frac{\delta _{l}N^{*I}}{\delta
\vartheta }\left( \left. \frac{\delta _{l}S_{N}}{\delta N^{I}}\right|
_{N=N^{*}}-L_{I}\right) =\left. \frac{\delta _{l}S_{N}}{\delta \vartheta }%
\right| _{N=N^{*}},  \label{buno}
\end{equation}
while for $\vartheta =N$ (\ref{fe}) gives 
\begin{equation}
\frac{\delta _{l}S_{L}}{\delta N^{I}}=L_{I}=\left. \frac{\delta _{l}S_{N}}{%
\delta N^{I}}\right| _{N=N^{*}}.  \label{bdue}
\end{equation}
Thus, $\left. \lfloor S_{N},S_{N}\rfloor \right| _{N=N^{*}}=\lfloor
S_{L},S_{L}\rfloor =0$. Moreover, 
\begin{equation}
\lfloor N^{I}-N^{*I},S_{L}\rfloor =0.  \label{billo}
\end{equation}
Indeed, differentiating $\lfloor S_{N},S_{N}\rfloor =0$ with respect to $%
N^{I}$ we get $\lfloor \delta _{l}S_{N}/\delta N^{I},S_{N}\rfloor =0$.
Setting $N=N^{*}$ in this equation and using 
\[
0=\frac{\delta _{r}L_{I}}{\delta \vartheta }=\frac{\delta _{r}}{\delta
\vartheta }\left( \left. \frac{\delta _{l}S_{N}}{\delta N^{I}}\right|
_{N=N^{*}}\right) =\left. \frac{\delta _{r}\delta _{l}S_{N}}{\delta
\vartheta \delta N^{I}}\right| _{N=N^{*}}+\left. \frac{\delta _{r}\delta
_{l}S_{N}}{\delta N^{J}\delta N^{I}}\right| _{N=N^{*}}\frac{\delta
_{r}(N^{*J}-N^{J})}{\delta \vartheta }, 
\]
where $\vartheta =\Phi ,K,N$ or $H$, we get 
\[
\left. \frac{\delta _{r}\delta _{l}S_{N}}{\delta N^{J}\delta N^{I}}\right|
_{N=N^{*}}\lfloor N^{J}-N^{*J},S_{L}\rfloor =0. 
\]
Removing the matrix of second derivatives we obtain (\ref{billo}).

For example, applying the inverse procedure just described to (\ref{gullo3})
we reconstruct (\ref{sln0}) and find that $S_{H}$ is given by (\ref{gullo2}).

\section{Proper canonical transformations}

\setcounter{equation}{0}

In this section we study the most general canonical transformations we need
to work with. In ref. \cite{masterf} it was shown that the master functional
allows us to describe the most general changes of field variables as linear
redefinitions of the form 
\begin{equation}
\Phi ^{A\hspace{0.01in}\prime }=\Phi ^{A}+N^{I}b_{I}^{A},\qquad N^{I\hspace{%
0.01in}\prime }=N^{J}z_{J}^{I},  \label{buah}
\end{equation}
where $b_{I}^{A}$ and $z_{J}^{I}$ are constants. We understand that these
constants can be both $c$-numbers and Grassmann variables, so from now on we
pay attention to their position. By convention, we place them to the right
of the proper fields and to the left of their sources.

Basically, the redefinitions (\ref{buah}) work as follows. Writing 
\begin{equation}
\Phi ^{A\hspace{0.01in}\prime }=\Phi ^{A}+\mathcal{O}_{\mathrm{R}}^{I}(\Phi
)b_{I}^{A}+\tilde{N}^{I}b_{I}^{A},  \label{bisu}
\end{equation}
we see that when we set $\tilde{N}^{I}=K=H=0$, which is supposed to switch
off the sectors of composite fields and gauge transformations, the change of
variables is practically $\Phi ^{A\hspace{0.01in}\prime }=\Phi ^{A}+\mathcal{%
O}_{\mathrm{R}}^{I}(\Phi )b_{I}^{A}$, which is the most general perturbative
field redefinition. However, the conditions $\tilde{N}^{I}=0$ switch off the
composite-field sector before the transformation, not after. Indeed, due to
the term $\tilde{N}^{I}b_{I}^{A}$ in (\ref{bisu}) after the transformation
the solutions of $L_{I}^{\prime }=\delta _{l}S_{N}^{\prime }/\delta
N^{I\prime \hspace{0.01in}}=0$ at $K=H=0$ are no longer $\tilde{N}^{I}=0$,
but some new $\tilde{N}^{I\hspace{0.01in}\prime }=0 $. Working out $\tilde{N}%
^{I\hspace{0.01in}\prime }$ it is found that at $\tilde{N}^{I\hspace{0.01in}%
\prime }=K=H=0$ the effective change of variables is corrected by $\mathcal{O%
}(b^{2})$-terms and finally reads 
\[
\Phi ^{A\hspace{0.01in}\prime }=\Phi ^{A}+\mathcal{O}_{\mathrm{R}}^{I}(\Phi )%
\tilde{b}_{I}^{A}, 
\]
where $\tilde{b}_{I}^{A}=b_{I}^{A}+\mathcal{O}(b^{2})$ is some calculable
series in powers of $b$. More details can be found in ref. \cite{masterf}.

Now, recall that the most general change of gauge-fixing can be obtained
shifting the sources $H_{I}$ by constants. We can embed (\ref{buah}) and the
change of gauge-fixing into the canonical transformation generated by 
\begin{equation}
F(\Phi ,K^{\prime },N,H^{\prime })=\int (\Phi
^{A}+N^{I}b_{I}^{A})K_{A}^{\prime }+\int N^{I}z_{I}^{J}(H_{J}^{\prime }-\xi
_{J}),  \label{fcann}
\end{equation}
which we call proper canonical transformation, where the $\xi _{I}$s are the
gauge-fixing parameters. More explicitly, we have 
\begin{eqnarray}
\Phi ^{A\hspace{0.01in}\prime } &=&\frac{\delta F}{\delta K_{A}^{\prime }}%
=\Phi ^{A}+N^{I}b_{I}^{A},\qquad K_{A}=\frac{\delta F}{\delta \Phi ^{A}}%
=K_{A}^{\prime },  \nonumber \\
N^{I\hspace{0.01in}\prime } &=&\frac{\delta F}{\delta H_{I}^{\prime }}%
=N^{J}z_{J}^{I},\qquad H_{I}=\frac{\delta F}{\delta N^{I}}%
=z_{I}^{J}(H_{J}^{\prime }-\xi _{J})+b_{I}^{A}K_{A}^{\prime }.  \label{buah2}
\end{eqnarray}
The proper action behaves as a scalar: 
\begin{equation}
S_{N}^{\prime }(\Phi ^{\prime },K^{\prime },N^{\prime },H^{\prime
})=S_{N}(\Phi ,K,N,H).  \label{transfop}
\end{equation}

Clearly, composing two proper canonical transformations we obtain a proper
canonical transformation.

Proper canonical transformations encode all transformations we need, because
they incorporate the most general field redefinitions and the most general
changes of gauge-fixing. Moreover, they can be used as true changes of field
variables, instead of mere replacements, because the external sources $K$
and $H$ transform without involving integrated fields. Finally, the
transformations are linear in $\Phi $ and $N$. This is important because the
Legendre transform is covariant only with respect to linear field
redefinitions.

The transformation (\ref{buah}) acts on the proper action and inside the
functional integral. It is easy to derive how it reflects on the generating
functionals. Inside $Z$ and $W$ it corresponds to the $J$-$L$-redefinitions 
\begin{equation}
J_{A}^{\prime }=J_{A},\qquad L_{I}^{\prime
}=(z^{-1})_{I}^{J}(L_{J}-b_{J}^{A}J_{A}^{\prime }).  \label{buaa}
\end{equation}
Indeed, we obtain 
\begin{equation}
\int \left( \Phi ^{A\hspace{0.01in}\prime }J_{A}^{\prime }+N^{I\hspace{0.01in%
}\prime }L_{I}^{\prime }\right) =\int \left( \Phi
^{A}J_{A}+N^{I}L_{I}\right) ,  \label{buana}
\end{equation}
so, because of (\ref{transfop}), the $Z$- and $W$-functionals behave as
scalars: 
\begin{equation}
Z^{\prime }(J^{\prime },K^{\prime },L^{\prime },H^{\prime
})=Z(J,K,L,H),\qquad W^{\prime }(J^{\prime },K^{\prime },L^{\prime
},H^{\prime })=W(J,K,L,H).  \label{agua}
\end{equation}
Formula (\ref{buana}), together with (\ref{transfop}), proves that a proper
canonical transformation manifestly preserves the conventional form of the
functional integral, which ensures that replacements and true changes of
field variables are practically the same thing.

It is also simple to see how (\ref{buah2}) reflects on $\Omega $. If we use (%
\ref{buaa}) inside (\ref{agua}) and the definition of master functional, we
find that the transformation (\ref{buah2}) looks exactly the same inside $%
\Omega $, as expected from its linearity. Moreover, $\Omega $ also behaves
as a scalar: $\Omega ^{\prime }(\Phi ^{\prime },K^{\prime },N^{\prime
},H^{\prime })=\Omega (\Phi ,K,N,H)$.

\section{Basic renormalization algorithm}

\label{basiccan}

\setcounter{equation}{0}

The master functional $\Omega $ allows us to work in a manifestly
field-covariant framework and, among the other things, describe
renormalization as a combination of parameter redefinitions and proper
canonical transformations, which are both operations that manifestly
preserve the master equation. To prove these statements, we proceed
gradually. In this section we a give a basic subtraction algorithm. In the
next section we generalize known cohomological theorems to the master
functional and use them to subtract divergences by means of generic (namely
not necessarily proper) canonical transformations and parameter
redefinitions. In section 7 we reach our final goal, namely show how to
extend the action so that all divergences can be subtracted by means of
parameter redefinitions and proper canonical transformations.

When it is necessary to specify a regularization technique, we use the
dimensional one. We assume that the proper classical action $S_{cN}$
satisfies the master equation $\lfloor S_{cN},S_{cN}\rfloor =0$ at the
regularized level, which ensures that no gauge anomalies are generated. Then
the renormalized master functional satisfies the proper master equation.

In this section we give the basic proof of renormalizability, where
divergences are subtracted just ``as they come'', using the minimal
subtraction scheme and the dimensional-regularization technique. We proceed
by replacements of actions, rathen than changes of variables. More
precisely, we replace the action with the action plus its counterterms, and
do not care to check if the counterterms can be incorporated inside the
action by means of parameter-redefinitions and canonical transformations.
This subtraction procedure does not even preserve the master equation at
each step. Nevertheless, it does preserve the master equation up to higher
orders, which is enough to ensure that the renormalized action obtained at
the end satisfies the master equation exactly.

Call $S_{N\hspace{0.01in}n}$ and $\Omega _{n}$ the proper action and the
master functional renormalized up to $n$ loops. Since we use the minimal
subtraction scheme, $S_{N\hspace{0.01in}n}=S_{cN}+$poles in $\varepsilon
=4-D $, where $D$ is the continued dimension. Moreover, we inductively
assume that $S_{N\hspace{0.01in}n}$ satisfies the proper master equation up
to higher orders, namely 
\begin{equation}
\lfloor S_{N\hspace{0.01in}n},S_{N\hspace{0.01in}n}\rfloor =\mathcal{O}%
(\hbar ^{n+1}).  \label{ostro}
\end{equation}
Formula (\ref{aididin}) implies $\lfloor \Omega _{n},\Omega _{n}\rfloor
=\langle \lfloor S_{N\hspace{0.01in}n},S_{N\hspace{0.01in}n}\rfloor \rangle =%
\mathcal{O}(\hbar ^{n+1})$. Now, $\lfloor S_{N\hspace{0.01in}n},S_{N\hspace{%
0.01in}n}\rfloor $ is a local functional, and $\langle \lfloor S_{N\hspace{%
0.01in}n},S_{N\hspace{0.01in}n}\rfloor \rangle $ is the functional that
collects the one-particle irreducible correlations functions containing one
insertion of $\lfloor S_{N\hspace{0.01in}n},S_{N\hspace{0.01in}n}\rfloor $.
Because of (\ref{ostro}), the $\mathcal{O}(\hbar ^{n+1})$-contributions to $%
\langle \lfloor S_{N\hspace{0.01in}n},S_{N\hspace{0.01in}n}\rfloor \rangle $
coincide with the $\mathcal{O}(\hbar ^{n+1})$-contributions to $\lfloor S_{N%
\hspace{0.01in}n},S_{N\hspace{0.01in}n}\rfloor $. Moreover, since $S_{N%
\hspace{0.01in}n}=S_{cN}+$poles and $\lfloor S_{cN},S_{cN}\rfloor =0$, we
have $\lfloor S_{N\hspace{0.01in}n},S_{N\hspace{0.01in}n}\rfloor =$poles.

Call $\Omega _{n\hspace{0.01in}\text{div}}^{(n+1)}$ the order-$(n+1)$
divergent part of $\Omega _{n}$. By the theorem of locality of counterterms,
it is a local functional, since all subdivergences have been subtracted
away. By the observations just made, and recalling that the classical limit
of $\Omega _{n}$ is $S_{cN}$, if we take the order-$(n+1)$ divergent part of 
$\lfloor \Omega _{n},\Omega _{n}\rfloor =\langle \lfloor S_{N\hspace{0.01in}%
n},S_{N\hspace{0.01in}n}\rfloor \rangle $, we get 
\begin{equation}
\lfloor S_{cN},\Omega _{n\hspace{0.01in}\text{div}}^{(n+1)}\rfloor =\frac{1}{%
2}\lfloor S_{N\hspace{0.01in}n},S_{N\hspace{0.01in}n}\rfloor +\mathcal{O}%
(\hbar ^{n+2}).  \label{cohoold}
\end{equation}
Now, if we define 
\begin{equation}
S_{N\hspace{0.01in}n+1}=S_{N\hspace{0.01in}n}-\Omega _{n\hspace{0.01in}\text{%
div}}^{(n+1)},  \label{buo}
\end{equation}
we still have $S_{N\hspace{0.01in}n+1}=S_{cN}+$poles, and, thanks to (\ref
{cohoold}), 
\[
\lfloor S_{N\hspace{0.01in}n+1},S_{N\hspace{0.01in}n+1}\rfloor =\mathcal{O}%
(\hbar ^{n+2}), 
\]
which promotes the inductive assumption to $n+1$ loops. Iterating the
argument, we are able to construct the renormalized proper action $S_{N%
\hspace{0.01in}}\equiv S_{N\hspace{0.01in}\infty }$ and the renormalized
master functional $\Omega _{\infty }$, and prove that both satisfy their
master equations exactly.

\section{Proper cohomology}

\setcounter{equation}{0}

The proof of renormalizability given in the previous section is
straightforward, but not completely satisfactory, because it does not take
advantage of the new formalism. The master equation is not preserved exactly
at each step, but only up to higher orders. Going through the proof it is
easy to convince oneself that it does not admit an immediate generalization
to arbitrary subtraction schemes. Moreover, since divergences are subtracted
as they come by the plain subtraction (\ref{buo}), we cannot even guarantee
that they can be reabsorbed into redefinitions of parameters and fields. We
are not even sure that the classical action $S_{cN}$ does contain enough
parameters to reabsorb divergences that way.

When we want to preserve the master equation exactly at every step of the
subtraction algorithm, the right-hand side of (\ref{ostro}) is zero, so $%
\lfloor \Omega _{n},\Omega _{n}\rfloor =0$. In the minimal subtraction
scheme, the $(n+1)$-loop divergent part of this equation gives the
cohomological problem 
\begin{equation}
\lfloor S_{cN},\Omega _{n\hspace{0.01in}\text{div}}^{(n+1)}\rfloor =0,
\label{chopr}
\end{equation}
instead of (\ref{cohoold}).

In this section we characterize the solutions of (\ref{chopr}) in a variety
of theories and use the results to upgrade the renormalization algorithm.
Let us first recall how we proceed in the usual formalism, by which we mean
the formalism where the antiparentheses are (\ref{usa}), we take (\ref{buh})
as the classical action $S(\Phi ,K)$, solution of the master equation $%
(S,S)=0$, we work with the generating functional $\Gamma (\Phi ,K)$ and in
all non-trivial cases describe renormalization as a replacement instead of a
true change of variables, plus redefinitions of parameters. In this context
the cohomological problem (\ref{chopr}) becomes 
\begin{equation}
(S,\Gamma _{n\hspace{0.01in}\text{div}}^{(n+1)})=0,  \label{cohopr2}
\end{equation}
where $\Gamma _{n\hspace{0.01in}\text{div}}^{(n+1)}$ is the $(n+1)$-loop
divergent part of the $n$-loop renormalized $\Gamma $-functional $\Gamma
_{n} $.

Let us assume that the most general solution of $(S,\chi )=0$, where $\chi $
is a local functional of zero ghost number and bosonic statistics, has the
form 
\begin{equation}
\chi (\Phi ,K)=\mathcal{G}(\Phi )+(S,\chi ^{\prime }),  \label{autoassu}
\end{equation}
where $\mathcal{G}$ and $\chi ^{\prime }$ are local functionals. The key
content of (\ref{autoassu}) is that the cohomologically non-trivial part $%
\mathcal{G}$ depends only on the fields $\Phi $. In a variety of common
cases, we can characterize $\mathcal{G}(\Phi )$ even more precisely.
Assuming that the set of fields $\Phi ^{A}$ is made of the classical fields $%
\phi $, the ghosts $C$, plus the gauge-trivial subsystem $\bar{C}$-$B$, then
using the theorem \ref{cohotheo} recalled in the appendix $\mathcal{G}(\Phi
) $ can be further decomposed as 
\begin{equation}
\mathcal{G}(\Phi )=\mathcal{G}^{\prime }(\phi )+(S,\chi ^{\prime \prime }),
\label{galassu}
\end{equation}
where $\mathcal{G}^{\prime }$ and $\chi ^{\prime \prime }$ are also local
functionals. This formula shows that the cohomologically non-trivial
solutions $\mathcal{G}^{\prime }(\phi )$ are the gauge-invariant terms
constructed with the classical fields $\phi $ and their covariant
derivatives.

For the arguments that follow we do not strictly need (\ref{galassu}), since
assumption (\ref{autoassu}) is sufficient. Once we have (\ref{autoassu}),
the renormalization of the $\Gamma $-functional proceeds as follows. Let $\{%
\mathcal{G}_{i}(\Phi )\}$ denote a basis for the non-trivial solutions $%
\mathcal{G}(\Phi )$ appearing in (\ref{autoassu}). Assume that the action $%
\mathcal{S}(\Phi )$ is a linear combination of all $\mathcal{G}_{i}(\Phi )$%
s, multiplied by independent parameters $\lambda _{i}$%
\begin{equation}
\mathcal{S}(\Phi )=\sum_{i}\lambda _{i}\mathcal{G}_{i}(\Phi ).  \label{exto}
\end{equation}
The action $S(\Phi ,K)$ still solves the master equation $(S,S)=0$, because $%
(S,\mathcal{G}_{i})=0$ and $(\mathcal{G}_{i},\mathcal{G}_{j})=0$. Note that
the identity $(\mathcal{G}_{i},\mathcal{G}_{j})=0$ is guaranteed by the fact
that the solutions $\mathcal{G}_{i}$ of the cohomological problem depend
only on the fields, and not on the sources $K$. If $\mathcal{G}_{i}$
depended on $K$, we would not be able to extend the classical action in such
a simple way, preserving the master equation. This is a key content of the
cohomological assumption (\ref{autoassu}).

Since $\Gamma _{n\hspace{0.01in}\text{div}}^{(n+1)}$ satisfies (\ref{cohopr2}%
), by the assumption (\ref{autoassu}) we can decompose it as 
\[
\Gamma _{n\hspace{0.01in}\text{div}}^{(n+1)}=\sum_{i}(\Delta _{n+1}\lambda
_{i})\mathcal{G}_{i}(\Phi )+(S,\chi _{n+1}^{\prime }), 
\]
where $\Delta _{n+1}\lambda _{i}$ are constants and $\chi _{n+1}^{\prime }$
is a local functional. The cohomologically trivial divergences $(S,\chi
_{n+1}^{\prime })$ can be subtracted by means of the canonical
transformation generated by 
\[
F_{n+1}(\Phi ,K^{\prime })=\int \Phi ^{A}K_{A}^{\prime }-\chi _{n+1}^{\prime
}(\Phi ,K^{\prime }), 
\]
while the divergences $\mathcal{G}_{i}(\Phi )$ are subtracted redefining the
parameters $\lambda _{i}$ as $\lambda _{i}^{\prime }=\lambda _{i}-\Delta
_{n+1}\lambda _{i}$. Indeed, if $S_{n}$ denotes the action renormalized up
to $n$ loops, we have 
\[
\Phi ^{\prime }=\Phi ^{A}-\frac{\delta \chi _{n+1}^{\prime }}{\delta K_{A}}%
,\qquad K^{\prime }=K+\frac{\delta \chi _{n+1}^{\prime }}{\delta \Phi }%
,\qquad S_{n}(\Phi ^{\prime },K^{\prime })=S_{n}(\Phi ,K)-(S,\chi
_{n+1}^{\prime }), 
\]
plus higher orders. Then we can define the $(n+1)$-loop renormalized action
as 
\[
S_{n+1}(\Phi ,K,\lambda )=S_{n}(\Phi ^{\prime },K^{\prime },\lambda ^{\prime
}) 
\]
and obtain 
\[
S_{n+1}(\Phi ,K,\lambda )=S_{n}(\Phi ,K,\lambda )-\sum_{i}(\Delta
_{n+1}\lambda _{i})\mathcal{G}_{i}(\Phi )-(S,\chi _{n+1}^{\prime
})=S_{n}(\Phi ,K,\lambda )-\Gamma _{n\hspace{0.01in}\text{div}}^{(n+1)} 
\]
plus higher orders.

In this procedure we started from the most general action (\ref{exto}),
containing all gauge invariant couplings $\lambda _{i}$. This option is the
safest one, from the point of view of renormalization, but then the theory
may contain more parameters than we actually need. For example, it may be
problematic to recognize finite theories and theories that are
renormalizable with a finite number of physical parameters. A possible
shortcut is to start from a lucky solution $\mathcal{S}(\Phi )$ and prove
that it does not get extended (as in the case of power-counting
renormalizable theories). If this is not possible, we can extend the action
step-by-step adding the missing $\lambda _{i}$s only when necessary.

Theorems that ensure (\ref{autoassu}) have been proved both for Yang-Mills
theory and gravity, for local composite fields and local functionals of
arbitrary ghost numbers \cite{coho}. In the rest of this section we prove
that once (\ref{autoassu}) holds in the usual formalism, it can be
straightforwardly promoted to the proper formalism. Then we show how to use
its generalized version in the proof of renormalizability.

We can start with the proper action (\ref{gullo3}), because we will show
that it just needs straightforward extensions, thanks to assumption (\ref
{autoassu}).

\begin{theorem}
If assumption (\ref{autoassu}) holds, the most general solution of the
cohomological problem $\lfloor S_{N\hspace{0.01in}},\chi \rfloor =0$, where $%
\chi (\Phi ,K,N,H)$ is a local functional, is 
\begin{equation}
\chi =\omega (\Phi )+\lfloor S_{N\hspace{0.01in}},\chi ^{\prime }\rfloor ,
\label{altra}
\end{equation}
where $\omega (\Phi )$ and $\chi ^{\prime }(\Phi ,K,N,H)$ are local
functionals.
\end{theorem}

\textit{Proof}. It is convenient to apply the procedures explained in
section 3 that allow us to switch back and forth between $S_{N}$ and $S_{L}$%
. Indeed, the subsystem $H$-$N$-$L$ has cohomological properties similar to
those of the common subsystem$\ \bar{C}$-$K_{\bar{C}}$-$B$ made of
antighosts, their sources and Lagrange multipliers, the sources associated
with $B$ and $L$ being missing. Integrating $L$ away, as we did in (\ref
{expsn}), is very much like integrating the Lagrange multipliers $B$ away.
Referring to the definitions given in the appendix, the subsystem $H$-$L$ is
gauge-semitrivial, because $\lfloor S_{L\hspace{0.01in}},H_{I}\rfloor
=(-1)^{\varepsilon _{I}}L_{I}$ and $\lfloor S_{L\hspace{0.01in}%
},L_{I}\rfloor =0$, while the subsystem $\bar{C}$-$B$ is usually
gauge-trivial. In the appendix we prove a theorem about the cohomological
properties of gauge-semitrivial subsystems, which we need to use here.

Define $\chi _{L}(\Phi ,K,H,L)=\left. \chi \right| _{N=N^{*}}$, where $%
N=N^{*}$ is the solution of (\ref{fe}). We have 
\[
\frac{\delta _{l}\chi _{L}}{\delta \vartheta }=\frac{\delta _{l}}{\delta
\vartheta }\left( \left. \chi \right| _{N=N^{*}}\right) =\left. \frac{\delta
_{l}\chi }{\delta \vartheta }\right| _{N=N^{*}}+\frac{\delta
_{l}(N^{*I}-N^{I})}{\delta \vartheta }\left. \frac{\delta _{l}\chi }{\delta
N^{I}}\right| _{N=N^{*}}, 
\]
where $\vartheta =\Phi ,K,N$ or $H$. Using (\ref{slll}) and (\ref{buno}-\ref
{billo}) we find 
\[
\lfloor S_{L\hspace{0.01in}},\chi _{L}\rfloor =\left. \lfloor S_{N\hspace{%
0.01in}},\chi \rfloor \right| _{N=N^{*}}-\lfloor S_{L},N^{I}-N^{*I}\rfloor
\left. \frac{\delta _{l}\chi }{\delta N^{I}}\right| _{N=N^{*}}=0. 
\]
Now, write 
\[
\chi _{L}(\Phi ,K,H,L)=f(\Phi ,K)+\mathcal{O}(L\text{-}H). 
\]
Using (\ref{gullo2}) and (\ref{sln0}) it is easy to check that 
\[
\lfloor S_{L\hspace{0.01in}},\chi _{L}\rfloor =(S,f)+\mathcal{O}(L\text{-}%
H), 
\]
where $S(\Phi ,K)$ is the solution (\ref{buh}) of the master equation $%
(S,S)=0$ when composite fields are switched off. Thus, $\lfloor S_{L\hspace{%
0.01in}},\chi _{L}\rfloor =0$ implies $(S,f)=0$. By assumption (\ref
{autoassu}) we know that the most general solution of this cohomological
condition is 
\[
f(\Phi ,K)=\omega (\Phi )+(S,\eta (\Phi ,K)), 
\]
where $\omega $ and $\eta $ are local functionals and $(S,\omega )=0$. It is
also easy to verify that we can write 
\[
f(\Phi ,K)=\omega (\Phi )+\lfloor S_{L\hspace{0.01in}},\eta \rfloor +%
\mathcal{O}(L\text{-}H) 
\]
and $\lfloor S_{L\hspace{0.01in}},\omega \rfloor =0$. Now, write 
\[
\chi _{L}(\Phi ,K,H,L)=\omega (\Phi )+\lfloor S_{L\hspace{0.01in}},\eta
\rfloor +\Delta \chi (\Phi ,K,H,L), 
\]
where $\lfloor S_{L},\Delta \chi \rfloor =0$ and $\Delta \chi =\mathcal{O}(L$%
-$H)$. Applying theorem \ref{theo9} we can conclude that there exists a
local functional $\chi _{L}^{\prime }(\Phi ,K,H,L)$ such that 
\begin{equation}
\chi _{L}=\omega (\Phi )+\lfloor S_{L\hspace{0.01in}},\chi _{L}^{\prime
}\rfloor .  \label{cohocoho}
\end{equation}
Finally, setting $L=L^{*}$ to switch back to $\chi $ and using (\ref{cucu}),
we obtain 
\begin{eqnarray*}
\chi (\Phi ,K,N,H) &=&\left. \chi _{L}\right| _{L=L^{*}}=\omega (\Phi
)+\left. \lfloor S_{L\hspace{0.01in}},\chi _{L}^{\prime }\rfloor \right|
_{L=L^{*}}= \\
&=&\omega (\Phi )+\lfloor S_{N\hspace{0.01in}},\left. \chi _{L}^{\prime
}\right| _{L=L^{*}}\rfloor -\lfloor S_{N},L_{I}^{*}\rfloor \left. \frac{%
\delta _{l}\chi _{L}^{\prime }}{\delta L_{I}}\right| _{L=L^{*}}=\omega (\Phi
)+\lfloor S_{N\hspace{0.01in}},\left. \chi _{L}^{\prime }\right|
_{L=L^{*}}\rfloor ,
\end{eqnarray*}
which is (\ref{altra}) with $\chi ^{\prime }=\left. \chi _{L}^{\prime
}\right| _{L=L^{*}}$.

\bigskip

Now renormalization can proceed as in the usual formalism. Assume that the
classical action $S_{N}$ is (\ref{gullo3}) with $\mathcal{S}(\Phi )$ given
by (\ref{exto}) and that the divergences are removed up to $n$ loops by
means of $\lambda $-redefinitions and canonical transformations. Denote the $%
n$-loop renormalized action and master functional with $S_{Nn}$ and $\Omega
_{n}$, respectively. Clearly, $\lfloor S_{Nn},S_{Nn}\rfloor =\lfloor \Omega
_{n},\Omega _{n}\rfloor =0$. We want to show that the $(n+1)$-loop
divergences can be removed in the same way. Call $\Omega _{n\hspace{0.01in}%
\text{div}}^{(n+1)}$ the $(n+1)$-loop divergent part of $\Omega _{n}$. By
the theorem of locality of counterterms, $\Omega _{n\text{div}}^{(n+1)}$ is
a local functional and, by the usual argument, $\lfloor \Omega _{n},\Omega
_{n}\rfloor =0$ implies 
\[
\lfloor S_{N},\Omega _{n\hspace{0.01in}\text{div}}^{(n+1)}\rfloor =0. 
\]
Applying the theorem proved above, write 
\begin{equation}
\Omega _{n\hspace{0.01in}\text{div}}^{(n+1)}=\sum_{i}(\Delta _{n+1}\lambda
_{i})\mathcal{G}_{i}(\Phi )+\lfloor S_{N},\chi _{n+1}^{\prime }\rfloor ,
\label{altra2}
\end{equation}
where $\chi _{n+1}^{\prime }$ is a local functional. We subtract the
divergent terms of type $\mathcal{G}_{i}$ redefining the parameters $\lambda
_{i}$ as $\lambda _{i}-\Delta _{n+1}\lambda _{i}$. The other divergent terms 
$\lfloor S_{N},\chi _{n+1}^{\prime }\rfloor $ are subtracted by means of a
canonical transformation generated by 
\begin{equation}
F_{n+1}(\Phi ,K^{\prime },N,H^{\prime })=\int \Phi ^{A}K_{A}^{\prime }+\int
N^{I}H_{I}^{\prime }-\chi _{n+1}^{\prime }(\Phi ,K^{\prime },N,H^{\prime }),
\label{fn+1}
\end{equation}
up to higher-orders, which, as usual, are dealt with at the subsequent steps
of the subtraction procedure.

This proves that when (\ref{autoassu}) holds and $\mathcal{S}(\Phi )$ is
general enough, renormalization can be achieved preserving the action (\ref
{gullo3}), up to canonical transformations. Yet, this procedure is still not
what we want. It is just the usual procedure generalized to the proper
formalism for the master functional, but it does not take advantage of the
new formalism. The canonical transformations (\ref{fn+1}) are not guaranteed
to be proper ones, so they are not covariant under the Legendre transform
and cannot be used as changes of variables inside the functional integral,
but just as replacements. In the next section we reorganize the subtraction
so as to achieve our goals.

\section{Renormalization of the master functional}

\label{RGren} \setcounter{equation}{0}

In this section we show how to extend the action (\ref{gullo3}) so that all
divergences proportional to the field equations and of gauge-fixing type are
subtracted making proper canonical transformations and all other divergences
are subtracted redefining parameters. These operations are fully covariant,
preserve the master equation and allow us to interpret the BR map, namely
the relation between bare and renormalized quantities, as a true change of
variables in the functional integral, instead of a mere replacement.

The results of this section apply to the most general gauge field theory
whose gauge algebra closes off shell and satisfies the cohomological
assumption (\ref{autoassu}), therefore also (\ref{altra}) in the proper
formalism for the master functional. We do not assume power-counting
renormalizability, nor that the number of parameters necessary to
renormalize divergences is finite. The search for new theories that are
renormalizable with a finite number of independent physical parameters, and
do not obey known criteria, is beyond the purposes of this paper.
Nevertheless, we do believe that the formalism developed here will help
organize that search in a more effective way.

For simplicity, we assume that the actions $S(\Phi ,K)$ and $S_{N}$ satisfy
their master equations at the regularized level, which guarantees that there
are no gauge anomalies. We can use any subtraction scheme that is compatible
with gauge invariance. When the master equation is not manifestly satisfied
at the regularized level we must check whether gauge anomalies cancel at one
loop or not. If they do, the Adler-Bardeen theorem \cite{adler} ensures that
there exists a subclass of subtraction schemes where they cancel to all
orders. Then our arguments work in that subclass of subtraction schemes.

Divergences proportional to the field equations and of gauge-fixing type can
actually be subtracted in two ways, which are good for different purposes: $%
i $) making (proper) canonical transformations; $ii$) redefining \textsl{%
ad-hoc} parameters, introduced just for that purpose. Option $i$) is
preferable if we want to check if our theory belongs to some special class
with respect to its renormalizability properties, for example it is finite
or renormalizable with a finite number of physical parameters. Option $ii$)
can be useful as a formal trick for intermediate purposes. It can be applied
trading wave-function renormalization constants for renormalization
constants of extra parameters put in front of kinetic and source terms.
However, when we use option $ii$) we may not realize that some divergent
terms can be subtracted by means of field redefinitions. This could lead us
to introduce parameters that may turn a finite or renormalizable theory into
a non-renormalizable one.

We describe how renormalization works with both options, beginning with
option $ii$) because it is a good introduction to option $i$).

\paragraph{Renormalization by redefinitions of parameters\newline
}

For the moment it is convenient to view renormalization as a redefinition of
parameters only, with no field redefinition, in the spirit of option $ii$).
We extend the classical action (\ref{gullo3}) so that it still satisfies the
master equation and contains enough independent parameters to subtract all
divergences by means of parameter redefinitions. Later we explain how to
implement option $i$).

The idea is as follows. The action (\ref{gullo3}) is not sufficiently
general to subtract all divergences by means of parameter redefinitions. To
say one thing, it is linear in $K$ and $H$, but it is easy to construct
divergent Feynman diagrams with more $K$- and $H$-external legs, in general.
As before, we can include all physical couplings $\lambda _{i}$ inside the
action $\mathcal{S}(\Phi )$, choosing the most general linear combination (%
\ref{exto}). Thus, the divergences of type $\mathcal{G}_{i}$ are subtracted
redefining the $\lambda _{i}$s and what remains are just cohomologically
trivial divergences $\lfloor S_{N\hspace{0.01in}},\chi _{n+1}^{\prime
}\rfloor $. Instead of subtracting them by means of canonical
transformations, we introduce \textsl{ad-hoc} parameters and redefine those.

We extend the action $S_{N}$ of (\ref{gullo3}) by means of the most general
perturbative (not necessarily proper) canonical transformation. Write its
generating function as 
\begin{equation}
F_{\subset }(\Phi ,K^{\prime },N,H^{\prime })=\int \Phi ^{A}K_{A}^{\prime
}+\int N^{I}H_{I}^{\prime }+\Delta F_{\subset }(\Phi ,K^{\prime
},N,H^{\prime }),  \label{finfty}
\end{equation}
where $\Delta F_{\subset }$ is the sum of all monomials we can construct
using the proper fields $\Phi $-$N$, the primed sources $K^{\prime }$-$%
H^{\prime }$ and their derivatives, multiplied by arbitrary independent
couplings $\zeta $. The transformation $F_{\subset }$ introduces many more
parameters than we actually need. For example, it includes enough parameters
to encode all field redefinitions. This happens just because we are applying
option $ii$). Clearly, this method is not the cleverest way to build a
covariant approach, and certainly it is not very practical. However, for the
moment we do so, and later explain how to economize on such redundancies
switching to option $i$).

With no loss of generality, we still use the minimal subtraction scheme for
the divergent parts. Once we have enough independent parameters $\lambda $, $%
\zeta $ to subtract all divergences by means of $\lambda $- and $\zeta $%
-redefinitions, it is straightforward to change to an arbitrary scheme
making finite $\lambda $- and $\zeta $-redefinitions.

We construct the extended classical action $S_{N\subset }$ composing the
action $S_{N}$ of (\ref{gullo3}) with the canonical transformation (\ref
{finfty}). We denote this operation as 
\begin{equation}
S_{N\subset }(\lambda ,\zeta )=S_{N}(\lambda )\circ F_{\subset }(\zeta ).
\label{snc}
\end{equation}
Clearly, $S_{N\subset }$ satisfies the proper master equation $\lfloor
S_{N\subset },S_{N\subset }\rfloor =0$.

If we use $S_{N\subset }$ as our starting classical action, renormalization
proceeds as follows. Assume, by induction, that the theory is renormalized
up to $n$ loops by means of $\lambda $- and $\zeta $-redefinitions. Call $%
S_{Nn\subset }(\lambda ,\zeta )$ and $\Omega _{n}$ the $n$-loop renormalized
proper action and master functional. Then the $(n+1)$-loop divergent part $%
\Omega _{n\hspace{0.01in}\text{div}}^{(n+1)}$ of the master functional is
local and satisfies the cohomological problem 
\[
\lfloor S_{N\subset },\Omega _{n\hspace{0.01in}\text{div}}^{(n+1)}\rfloor
=0. 
\]
Applying the inverse $F_{\subset }^{-1}$ of (\ref{finfty}) we obtain 
\[
\lfloor S_{N},\Omega _{n\hspace{0.01in}\text{div}}^{(n+1)}\circ F_{\subset
}^{-1}\rfloor =0, 
\]
therefore, using (\ref{altra}) and applying $F_{\subset }$ again, we get 
\begin{equation}
\Omega _{n\hspace{0.01in}\text{div}}^{(n+1)}=\sum_{i}(\Delta \lambda _{i})%
\mathcal{G}_{i}(\Phi )\circ F_{\subset }+\lfloor S_{N\subset },\chi
_{n+1\subset }^{\prime }\rfloor ,  \label{ibsu}
\end{equation}
where $\chi _{n+1\subset }^{\prime }$ is a local functional. The divergences
of type $\mathcal{G}$ can still be removed redefining the $\lambda $s.
Indeed, we have 
\[
S_{N\subset }(\lambda -\Delta \lambda ,\zeta )=S_{N\subset }(\lambda ,\zeta
)-\sum_{i}(\Delta \lambda _{i})\mathcal{G}_{i}(\Phi )\circ F_{\subset }%
\hspace{0.01in}. 
\]
Since $S_{Nn\subset }(\lambda ,\zeta )=S_{N\subset }(\lambda ,\zeta )+%
\mathcal{O}(\hbar )$, we also have 
\[
S_{Nn\subset }(\lambda -\Delta \lambda ,\zeta )=S_{Nn\subset }(\lambda
,\zeta )-\sum_{i}(\Delta \lambda _{i})\mathcal{G}_{i}(\Phi )\circ F_{\subset
}+\mathcal{O}(\hbar ^{n+2})\ . 
\]

Now, consider the cohomologically trivial divergences $\lfloor S_{N\subset
},\chi _{n+1\subset }^{\prime }\rfloor $. They can be subtracted by means of
the canonical transformation generated by 
\[
F_{-}(\Phi ,K^{\prime },N,H^{\prime })=\int \Phi ^{A}K_{A}^{\prime }+\int
N^{I}H_{I}^{\prime }-\chi _{n+1\subset }^{\prime }(\Phi ,K^{\prime
},N,H^{\prime }), 
\]
up to higher orders. Indeed, 
\[
S_{N\subset }(\lambda -\Delta \lambda ,\zeta )\circ F_{-}=S_{N\subset
}(\lambda ,\zeta )-\sum_{i}(\Delta \lambda _{i})\mathcal{G}_{i}(\Phi )\circ
F_{\subset }-\lfloor S_{N\subset },\chi _{n+1\subset }^{\prime }\rfloor
=S_{N\subset }(\lambda ,\zeta )-\Omega _{n\hspace{0.01in}\text{div}%
}^{(n+1)}, 
\]
up to higher orders. Now, 
\[
S_{N\subset }(\lambda -\Delta \lambda ,\zeta )\circ F_{-}=S_{N}(\lambda
-\Delta \lambda )\circ F_{\subset }(\zeta )\circ F_{-}. 
\]
Certainly, $F_{\subset }(\zeta )\circ F_{-}$ is a canonical transformation
equal to $F_{\subset }$ plus $(n+1)$-loop corrections. Since $F_{\subset
}(\zeta )$ contains all allowed perturbative terms multiplied by independent
parameters, there must exist redefinitions $\zeta \rightarrow \zeta -\Delta
\zeta $ such that $F_{\subset }(\zeta )\circ F_{-}=F_{\subset }(\zeta
-\Delta \zeta )$ up to higher orders. Thus, 
\[
S_{N\subset }(\lambda -\Delta \lambda ,\zeta -\Delta \zeta )=S_{N\subset
}(\lambda ,\zeta )-\Omega _{n\hspace{0.01in}\text{div}}^{(n+1)}+\mathcal{O}%
(\hbar ^{n+2}). 
\]

Finally, the $(n+1)$-loop renormalized action 
\[
S_{Nn+1\subset }(\lambda ,\zeta )=S_{Nn\subset }(\lambda -\Delta \lambda
,\zeta -\Delta \zeta ) 
\]
is equal to $S_{Nn\subset }(\lambda ,\zeta )-\Omega _{n\hspace{0.01in}\text{%
div}}^{(n+1)}+\mathcal{O}(\hbar ^{n+2})$, which proves that we can subtract
all divergences redefining the parameters $\lambda $ and $\zeta $.

\paragraph{Renormalization by redefinitions of parameters and proper
canonical transformations\newline
}

We have just shown how the proper action $S_{N}$ can be extended so that it
contains enough independent parameters to subtract all divergences by means
of parameter redefinitions. Now we explain how to proceed when we want,
instead, to subtract the divergences proportional to the field equations and
those of gauge-fixing type by means of proper canonical transformations. We
just need to restrict $F_{\subset }(\zeta )$ dropping all contributions that
are unnecessary. Since $\Delta F_{\subset }(\Phi ,K^{\prime },N,H^{\prime })$
is linear in the parameters $\zeta $, we first analyze it to the first order
in $\zeta $. Observe that $\chi _{n+1\subset }^{\prime }$ in (\ref{ibsu}) is
equivalent to $\chi _{n+1\subset }^{\prime }+\lfloor S_{N\subset },\eta
_{n+1\subset }\rfloor $, where $\eta _{n+1\subset }$ is an arbitrary local
functional. Correspondingly, the redefinitions of certain parameters $\zeta $
contained in $F_{\subset }$ have no effect on $\lfloor S_{N\subset },\chi
_{n+1\subset }^{\prime }\rfloor $ and $\Omega _{n\hspace{0.01in}\text{div}%
}^{(n+1)}$. We can use this freedom to simplify $\Delta F_{\subset }$.
Another freedom is of course to compose $F_{\subset }$ with proper canonical
transformations.

Using (\ref{fe}) and (\ref{arrarrarra}) recall that 
\begin{equation}
(-1)^{\varepsilon _{I}}\lfloor S_{N},H_{I}\rfloor =\frac{\delta _{l}S_{N}}{%
\delta N^{I}}=L_{I}=A_{IJ}\tilde{N}_{S}^{J}+\mathcal{O}(\tilde{N}_{S})\text{%
-perturbative corrections}.  \label{para}
\end{equation}
It is convenient to express the combinations $\tilde{N}_{S}$ as functions of 
$L_{I}$ and organize $\Delta F_{\subset }(\Phi ,K^{\prime },N,H^{\prime })$
as an expansion in powers of $K^{\prime },H^{\prime }$ and $L$. Clearly, $%
\lfloor S_{N},L_{I}\rfloor =0$. Write 
\begin{equation}
\Delta F_{\subset }(\Phi ,K^{\prime },N,H^{\prime })=f_{\subset }(\Phi
,K^{\prime },L,H^{\prime }).  \label{hus}
\end{equation}
The source-independent sector $f_{\subset }(\Phi ,0,L,0)$ collects the most
general changes of gauge-fixing. Write 
\[
f_{\subset }(\Phi ,0,L,0)=\Psi _{\subset }(\Phi )+\mathcal{O}(L). 
\]
The functional $\Psi _{\subset }(\Phi )$ can be dropped, because its
effects, which are the usual changes of gauge-fixing, can also be produced
by the constants $\xi _{I}$ contained in the proper canonical transformation
(\ref{fcann}). The $\mathcal{O}(L)$-corrections, which are also $\mathcal{O}(%
\tilde{N}_{S})$-corrections, are more general changes of gauge-fixing that
depend on the sources of composite fields. Using (\ref{para}) the $\mathcal{O%
}(L)$-terms can be written as 
\begin{equation}
\int L_{I}U^{I}(\Phi ,L)=\int (-1)^{\varepsilon _{I}}\lfloor
S_{N},H_{I}\rfloor U^{I}=\lfloor S_{N},\int (-1)^{\varepsilon
_{I}}H_{I}U^{I}\rfloor +\int H_{I}\lfloor S_{N},U^{I}\rfloor ,  \label{bua}
\end{equation}
where $U^{I}$ are local functions. Working to the first order in $\zeta $, $%
f_{\subset }$ and $U^{I}$ must be considered $\mathcal{O}(\zeta )$, so the
first term after the equals sign can be also written as 
\[
\lfloor S_{N\subset },\int (-1)^{\varepsilon _{I}}H_{I}U^{I}\rfloor . 
\]
For the reasons explained above, we drop it, and therefore remain with the
last term of (\ref{bua}), which is proportional to $H$. We conclude that we
can take $f_{\subset }(\Phi ,0,L,0)=0$.

Next, consider the terms of $f_{\subset }$ that are linear in $K^{\prime }$
and $H^{\prime }$. Expanding in the basis of composite fields, we can write
them as 
\begin{equation}
\int \mathcal{O}^{I}(\Phi )u_{I}^{A}K_{A}^{\prime }+\int \mathcal{O}%
^{I}(\Phi )v_{I}^{J}H_{J}^{\prime },  \label{bul}
\end{equation}
where $u_{I}^{A}$ and $v_{J}^{I}$ are constants. Composing $F_{\subset }$
with an infinitesimal proper canonical transformation (\ref{fcann}) with $%
b_{I}^{A}=-u_{I}^{A}$, $z_{I}^{J}=-v_{I}^{J}$ and $\xi _{I}=0$, the total
can be cast in the form 
\begin{equation}
-\int \tilde{N}_{S}^{I}u_{I}^{A}K_{A}^{\prime }-\int \tilde{N}%
_{S}^{I}v_{I}^{J}H_{J}^{\prime },  \label{bil}
\end{equation}
to the first order in $\zeta $. Since $\tilde{N}_{S}=\mathcal{O}(L)$, we
have been able to convert the terms (\ref{bul}), which are $\mathcal{O}%
(K^{\prime })$ and $\mathcal{O}(H^{\prime })$, into terms that are $\mathcal{%
O}(K^{\prime }L)$ and $\mathcal{O}(H^{\prime }L)$. This means that we can
drop all terms of type (\ref{bul}) from $f_{\subset }$.

Recapitulating, we take $f_{\subset }$ to be the most general sum of terms $%
\mathcal{O}(\tilde{K}^{\prime })\mathcal{O}(L)+\mathcal{O}(\tilde{K}^{\prime 
\hspace{0.01in}2})$, each monomial being multiplied by an independent
parameter $\zeta $. Here $\tilde{K}^{\prime }$ is a symbolic notation that
collects both $K^{\prime }$ and $H^{\prime }$. Similarly, let $\tilde{\Phi}$
collect $\Phi $ and $N$. Define $F_{\subset }$ from (\ref{finfty}) and (\ref
{hus}), where now $f_{\subset }$ is of restricted type. From now on we stop
focusing on the first order in $\zeta $, and our results apply to all orders
in $\zeta $. The canonical transformation $F_{\subset }$ produces a
field/source redefinition of the form 
\begin{equation}
\tilde{\Phi}^{\prime }=\tilde{\Phi}+\mathcal{O}(L)+\mathcal{O}(\tilde{K}%
^{\prime }),\qquad \tilde{K}=\tilde{K}^{\prime }+\mathcal{O}(\tilde{K}%
^{\prime }).  \label{bullo}
\end{equation}
Obviously, composing two canonical transformations of restricted type, we
get a canonical transformation of restricted type. Moreover, every canonical
transformation (\ref{bullo}) is generated by an $F_{\subset }$ of restricted
type.

Define the extended classical action $S_{N\subset }$ as in (\ref{snc}),
where now $F_{\subset }(\zeta )$ is of restricted type. Assume, by
induction, that the theory is renormalized by $\lambda $- and $\zeta $%
-redefinitions, plus proper canonical transformations, up to $n$ loops. Let $%
S_{Nn\subset }$ and $\Omega _{n}$ denote the $n$-loop renormalized proper
action and master functional. Consider the $(n+1)$-loop divergent terms $%
\Omega _{n\hspace{0.01in}\text{div}}^{(n+1)}$ of $\Omega _{n}$. Clearly,
they are still decomposed as in (\ref{ibsu}) and the divergences of type $%
\mathcal{G}_{i}(\Phi )\circ F_{\subset }$ can still be subtracted away
redefining the parameters $\lambda $ contained in $\mathcal{S}(\Phi )$. We
want to show that the divergences of type $\lfloor S_{N\subset },\chi
_{n+1\subset }^{\prime }\rfloor $ can be subtracted redefining the
parameters $\zeta $ of the restricted generating function $F_{\subset }$ and
making a proper canonical transformation.

Define $L_{I\subset }=L_{I}\circ F_{\subset }$ and $H_{I\subset }=H_{I}\circ
F_{\subset }$. Using (\ref{para}) and (\ref{bullo}) we have 
\begin{equation}
L_{I\subset }=(-1)^{\varepsilon _{I}}\lfloor S_{N\subset },H_{I\subset
}\rfloor =L_{I}+\mathcal{O}(L)+\mathcal{O}(\tilde{K}).  \label{gu}
\end{equation}
Consider $\chi _{n+1\subset }^{\prime }$ as a functional of $\Phi $, $%
L_{\subset }$, $K$ and $H$, and expand it in powers of $L_{\subset }$, $K$
and $H$. Expanding the contribution at $L_{\subset }=K=H=0$ in the basis of
composite fields, write 
\[
\chi _{n+1\subset }^{\prime }=\int \mathcal{O}^{I}(\Phi )q_{I}+\int
L_{I\subset }U^{I}(\Phi ,L_{\subset })+\mathcal{O}(\tilde{K}), 
\]
where $q_{I}$ are $\mathcal{O}(\hbar ^{n+1})$-constants and $U^{I}$ are
local functions. With a proper canonical transformation generated by 
\[
F_{q}(\Phi ,K^{\prime },N,H^{\prime })=\int \Phi ^{A}K_{A}^{\prime }+\int
N^{I}(H_{I}^{\prime }-q_{I}), 
\]
we can add a term $-\int N^{I}q_{I}$ to $\chi _{n+1\subset }^{\prime }$ and
then remain with 
\[
\chi _{n+1\subset }^{\prime \prime }=-\int \tilde{N}_{S}^{I}q_{I}+\int
L_{I\subset }U^{I}(\Phi ,L_{\subset })+\mathcal{O}(\tilde{K}). 
\]
Using (\ref{para}) and (\ref{gu}) we can write this formula as 
\[
\chi _{n+1\subset }^{\prime \prime }=\int L_{I\subset }V^{I}(\Phi
,L_{\subset })+\mathcal{O}(\tilde{K}), 
\]
for some new local functions $V^{I}$. Now, use (\ref{gu}) again to write 
\[
\chi _{n+1\subset }^{\prime \prime }=\int (-1)^{\varepsilon _{I}}\lfloor
S_{N\subset },H_{I\subset }\rfloor V^{I}+\mathcal{O}(\tilde{K})=\lfloor
S_{N\subset },\int (-1)^{\varepsilon _{I}}H_{I\subset }V^{I}\rfloor +\int
H_{I\subset }\lfloor S_{N\subset },V^{I}\rfloor +\mathcal{O}(\tilde{K}). 
\]
The first term after the equals sign can be dropped, since the divergences
we have to remove are $\lfloor S_{N\subset },\chi _{n+1\subset }^{\prime
\prime }\rfloor $. Thus we remain with $\chi _{n+1\subset }^{\prime \prime
\prime }=\mathcal{O}(\tilde{K})$. Expanding in the basis of composite
fields, write 
\[
\chi _{n+1\subset }^{\prime \prime \prime }=\int \mathcal{O}^{I}(\Phi
)m_{I}^{A}K_{A}+\int \mathcal{O}^{I}(\Phi )n_{I}^{J}H_{J}+\mathcal{O}(\tilde{%
K})\mathcal{O}(L_{\subset })+\mathcal{O}(\tilde{K}^{2}), 
\]
where $m_{I}^{A}$ and $n_{J}^{I}$ are $\mathcal{O}(\hbar ^{n+1})$-constants.
With an infinitesimal proper canonical transformation generated by 
\[
F_{mn}(\Phi ,K^{\prime },N,H^{\prime })=\int (\Phi
^{A}-N^{I}m_{I}^{A})K_{A}^{\prime }+\int N^{I}(H_{I}^{\prime
}-n_{I}^{J}H_{J}^{\prime }), 
\]
we can arrive at 
\[
\chi _{n+1\subset }^{\prime \prime \prime \prime }=-\int \tilde{N}%
_{S}^{I}m_{I}^{A}K_{A}-\int \tilde{N}_{S}^{I}n_{I}^{J}H_{J}+\mathcal{O}(%
\tilde{K})\mathcal{O}(L_{\subset })+\mathcal{O}(\tilde{K}^{2}). 
\]
Using (\ref{para}) and (\ref{gu}) we find 
\[
\chi _{n+1\subset }^{\prime \prime \prime \prime }=\mathcal{O}(\tilde{K})%
\mathcal{O}(L_{\subset })+\mathcal{O}(\tilde{K}^{2})=\mathcal{O}(\tilde{K})%
\mathcal{O}(L)+\mathcal{O}(\tilde{K}^{2}). 
\]
These terms are precisely of the type that can be subtracted away with a
restricted infinitesimal transformation. Let $F_{r}$ denote the generating
function of such a transformation. Collecting all operations made so far, we
have 
\[
S_{N\subset }(\lambda -\Delta \lambda ,\zeta )\circ F_{r}\circ F_{q}\circ
F_{mn}=S_{N\subset }(\lambda ,\zeta )-\Omega _{n\hspace{0.01in}\text{div}%
}^{(n+1)}+\mathcal{O}(\hbar ^{n+2}). 
\]
Now, write 
\[
S_{N\subset }(\lambda -\Delta \lambda ,\zeta )\circ F_{r}=S_{N}(\lambda
-\Delta \lambda )\circ F_{\subset }(\zeta )\circ F_{r}\ . 
\]
Since $F_{\subset }(\zeta )\circ F_{r}$ is the composition of two canonical
transformations of restricted type, it is of restricted type. Since $%
F_{\subset }$ is the most general transformation of restricted type, there
must exist $\zeta $-redefinitions $\zeta -\Delta \zeta $ such that $%
F_{\subset }(\zeta )\circ F_{r}=F_{\subset }(\zeta -\Delta \zeta )$.
Moreover, $F_{p}\equiv F_{q}\circ F_{mn}$ is the composition of two proper
canonical transformations, so it is a proper canonical transformation. Thus, 
\[
S_{N\subset }(\lambda -\Delta \lambda ,\zeta -\Delta \zeta )\circ
F_{p}=S_{N\subset }(\lambda ,\zeta )-\Omega _{n\hspace{0.01in}\text{div}%
}^{(n+1)}+\mathcal{O}(\hbar ^{n+2}). 
\]
Finally, define the $(n+1)$-loop renormalized action as 
\[
S_{Nn+1\subset }(\lambda ,\zeta )=S_{Nn\subset }(\lambda -\Delta \lambda
,\zeta -\Delta \zeta )\circ F_{p}\hspace{0.01in}. 
\]
Since $S_{Nn\subset }(\lambda ,\zeta )=S_{N\subset }(\lambda ,\zeta )+%
\mathcal{O}(\hbar )$, we also have 
\[
S_{Nn+1\subset }(\lambda ,\zeta )=S_{Nn\subset }(\lambda ,\zeta )-\Omega _{n%
\hspace{0.01in}\text{div}}^{(n+1)}+\mathcal{O}(\hbar ^{n+2}), 
\]
so the $(n+1)$-loop divergences are subtracted in the way we want.

We have thus achieved our goals: 1) all divergences proportional to the
field equations, as well as those of gauge-fixing type, are subtracted
making proper canonical transformations; 2) all divergences of other types
are subtracted by means of parameter redefinitions. Clearly, the master
equation is preserved at each step of the subtraction.

The form of the action (\ref{gullo3}) is not invariant under proper
canonical transformations, so it is worth to describe how to proceed when we
do not start from a classical proper action $S_{N}$ equal to (\ref{gullo3}),
but we take an action $S_{N}^{\prime }=S_{N}\circ F_{P}$, equal to (\ref
{gullo3}) composed with a convergent proper canonical transformation $F_{P}$%
. Then, by the results of section 4 the master functional $\Omega $ is
turned into $\Omega ^{\prime }=\Omega \circ F_{P}$. All arguments presented
above are modified just replacing the canonical transformations $F_{X}$ with 
$F_{X}^{\prime }=F_{P}^{-1}\circ F_{X}\circ F_{P}$. If $F_{X}$ is proper,
then $F_{X}^{\prime }$ is also proper. If $F_{X}$ is of restricted type,
then $F_{X}^{\prime }$ defines the canonical transformations of restricted
type for the starting action $S_{N}^{\prime }$.

So, for example, we first extend the classical action $S_{N}^{\prime }$ to $%
S_{N\hspace{0.01in}\subset }^{\prime }=S_{N}^{\prime }\circ F_{\subset
}^{\prime }=S_{N\hspace{0.01in}\subset }\circ F_{P}$. Then, we assume by
induction that the $n$-loop renormalized action $S_{N\hspace{0.01in}n\subset
}^{\prime }(\lambda ,\zeta )$ is obtained from $S_{N\hspace{0.01in}\subset
}^{\prime }(\lambda ,\zeta )$ redefining the parameters $\lambda $, $\zeta $
and making proper canonical transformations $F_{X}^{\prime }$. This means
that we can write $S_{N\hspace{0.01in}n\subset }^{\prime }(\lambda ,\zeta
)=S_{N\hspace{0.01in}n\subset }(\lambda ,\zeta )\circ F_{P}$, where $S_{N%
\hspace{0.01in}n\subset }(\lambda ,\zeta )$ is obtained from $S_{N\hspace{%
0.01in}\subset }(\lambda ,\zeta )$ also making parameter redefinitions and
proper canonical transformations $F_{X}$. Because of this assumption, we
have $\Omega _{n}^{\prime }=\Omega _{n}\circ F_{P}$, so the $(n+1)$-loop
divergent part of $\Omega _{n}^{\prime }$ is $\Omega _{n\hspace{0.01in}\text{%
div}}^{(n+1)\prime }$ $=\Omega _{n\hspace{0.01in}\text{div}}^{(n+1)}\circ
F_{P}$, and it can be subtracted defining 
\[
S_{N\hspace{0.01in}n+1\subset }^{\prime }(\lambda ,\zeta )=S_{N\hspace{0.01in%
}n\subset }(\lambda -\Delta \lambda ,\zeta -\Delta \zeta )\circ F_{p}\circ
F_{P}=S_{N\hspace{0.01in}n\subset }^{\prime }(\lambda -\Delta \lambda ,\zeta
-\Delta \zeta )\circ F_{p}^{\prime }. 
\]
This result shows that divergences can still be subtracted redefining the
parameters $\lambda $, $\zeta $ and making proper canonical transformations $%
F_{p}^{\prime }$.\ Thus, the inductive assumption is correctly promoted up
to $n+1$ loops, so it is also promoted to $n=\infty $. The final
renormalized master functional is just $\Omega _{\infty }^{\prime }=\Omega
_{\infty }\circ F_{P}$, which is obviously convergent.

\section{Conclusions}

\setcounter{equation}{0}

In this paper we have formulated a general field-covariant approach to
quantum gauge field theory, generalizing the ideas of ref.s \cite
{fieldcov,masterf}. Instead of working with the usual generating functional $%
\Gamma (\Phi ,K)$ of one-particle irreducible diagrams, we have introduced a
master functional $\Omega (\Phi ,K,N,H)$ that depends on the usual pair of
field/source conjugate variables $\Phi $ and $K$, plus a new pair $N$ and $H$%
, associated with the composite fields and their gauge transformations. The
functional $\Omega $ is also a generating functional of one-particle
irreducible diagrams and it is defined as the Legendre transform of the
improved functional $W(J,K,L,H)$ with respect to both sources $J$ and $L$ of
elementary and composite fields. We have extended the definitions of
antiparentheses, canonical transformations and the master equation to the
new formalism. The master functional satisfies the extended master equation.
In the proper approach the classical action $S_{N}(\Phi ,K,N,H)$ coincides
with the classical limit of $\Omega $. Moreover, both $\Phi $ and $N$ are
integrated fields, while $K$ and $H$ are external sources. The functional $%
\Omega $ collects the set of one-particle irreducible diagrams determined by 
$S_{N}$ with the usual Feynman rules.

The most general perturbative field redefinitions and changes of
gauge-fixing can be expressed by means of proper canonical transformations,
under which the action $S_{N}$ behaves as a scalar. Precisely, the proper
canonical transformations are linear, and such that the proper fields $\Phi $
and $N$ transform independently of the sources $K$ and $H$, and viceversa.
Thanks to this, we can implement the transformations as true changes of
variables in the functional integral, instead of using them as mere
replacements. This property overcomes an important difficulty of the old
approach, where it impossible to use canonical transformations as changes of
variables in the functional integral whenever the external sources $K$ are
transformed into non-trivial functions of $\Phi $. Being linear, proper
canonical transformations look identical inside the master functional $%
\Omega $, which also behaves as a scalar.

The approach of this paper is covariant with respect to the most general
perturbative field redefinitions and allows us to prove the
renormalizability of gauge theories in a general field-covariant setting.
When cohomological theorems hold, we have shown how to generalize them to
the proper formalism. When there are no gauge anomalies, we have shown that
the classical action $S_{N}$ can be extended so that it contains enough
parameters to subtract all divergences that are proportional to the field
equations and of gauge-fixing type by means of proper canonical
transformations and all other divergences by means of parameter
redefinitions. The master equation is exactly preserved at every step of the
subtraction procedure.

\vskip 25truept \noindent {\Large \textbf{Appendix\quad A cohomological
theorem for gauge semitrivial subsystems}}

\vskip 15truept

\renewcommand{\theequation}{A.\arabic{equation}} \setcounter{equation}{0}

In this appendix we prove cohomological properties that are useful to
classify gauge-invariant terms. The first theorem is a known result, the
second theorem is a new generalization, important for some applications
contained in the paper.

For definiteness, let $H$ and $U$ denote two fields of the set $\Phi $ and $%
S(\Phi ,K)$ denote the action, solution of the master equation $(S,S)=0$,
where $(.,.)$ are the usual antiparentheses (\ref{usa}). Write $\Phi =(\Phi
_{r},H,U)$. With obvious modifications everything we are going to say also
applies when the fields, the action and the antiparentheses are replaced by
the proper fields, the proper action and the squared antiparentheses.

We say that $H$ and $U$ form a \textit{gauge-semitrivial} $\Phi $-subsystem
if $(S,H)=U$, $(S,U)=0$. We say that $H$ and $U$ form a \textit{gauge-trivial%
} $\Phi $-subsystem if it is gauge-semitrivial and $(S,\Phi _{r})$ is
independent of $H$ and $U$. We call the $\Phi $-subset $\Phi _{r}$ \textit{%
gauge-irreducible} if it does not contain gauge-semitrivial subsystems.

For example, if $\Phi $ is made of the classical fields $\phi $, the ghosts $%
C$, the antighosts $\bar{C}$ and the Lagrange multipliers $B$, and the
action has the usual form 
\begin{equation}
S(\Phi ,K)=\mathcal{S}(\Phi )-\int R_{\phi }(\phi ,C)K_{\phi }-\int
R_{C}(C)K_{C}-\int B^{a}K_{\bar{C}}^{a},  \label{actr}
\end{equation}
then the $\Phi $-subsystem $\bar{C}$-$B$ is gauge-trivial, because $(S,\bar{C%
})=B$ and $(S,B)=0$, while $(S,\phi )$ and $(S,C)$ depend only on $\phi $
and $C$.

\begin{theorem}
Let $H$-$U$ be a gauge-trivial $\Phi $-subsystem and $\chi (\Phi )$ a local
functional such that $(S,\chi )=0$. Then there exist local functionals $%
f(\Phi _{r})$ and $\chi ^{\prime }(\Phi )$ such that 
\begin{equation}
\chi (\Phi )=f(\Phi _{r})+(S,\chi ^{\prime }).  \label{tesi}
\end{equation}
\label{cohotheo}
\end{theorem}

\textit{Proof}. Expand $\chi (\Phi )$ as 
\[
\chi (\Phi )=f(\Phi _{r})+\sum_{m+n>0}F_{m,n}, 
\]
where 
\[
F_{m,n}=\int f_{t_{1}\cdots t_{m},s_{1}\cdots s_{n}}(\Phi
_{r})U^{t_{1}}\cdots U^{t_{m}}H^{s_{1}}\cdots H^{s_{n}} 
\]
and $f_{t_{1}\cdots t_{m},s_{1}\cdots s_{n}}$ is a local operator-function,
by which we mean that it can contain derivatives acting on the $U$s and the $%
H$s. To simplify the notation, write 
\[
F_{m,n}^{\prime }\equiv \int (S,f_{t_{1}\cdots t_{m},s_{1}\cdots
s_{n}})U^{t_{1}}\cdots U^{t_{m}}H^{s_{1}}\cdots H^{s_{n}}. 
\]
Imposing $(S,\chi )=0$ we get $(S,f)=0$ and 
\begin{equation}
F_{m,n}^{\prime }+\int U^{s}\frac{\delta _{l}}{\delta H^{s}}F_{m-1,n+1}=0.
\label{acceu}
\end{equation}
Observe that 
\begin{equation}
\int U^{t}\frac{\delta _{l}}{\delta U^{t}}F_{m,n}=mF_{m,n},\qquad \int H^{s}%
\frac{\delta _{l}}{\delta H^{s}}F_{m,n}=nF_{m,n}.  \label{effo}
\end{equation}
Applying the operation 
\begin{equation}
\int H^{t}\frac{\delta _{l}}{\delta U^{t}}  \label{oppe}
\end{equation}
to equation (\ref{acceu}) and using (\ref{effo}), we get the useful identity 
\begin{equation}
0=\int H^{t}\frac{\delta _{l}}{\delta U^{t}}F_{m,n}^{\prime
}+(n+1)F_{m-1,n+1}+\int (-1)^{\varepsilon _{s}\varepsilon _{t}}H^{t}U^{s}%
\frac{\delta _{l}^{2}}{\delta U^{t}\delta H^{s}}F_{m-1,n+1},  \label{noval}
\end{equation}
where $\varepsilon _{s}$, $\varepsilon _{t}$ denote the statistics of $U^{s}$%
, $H^{t}$. With the help of this identity and (\ref{effo}) again we find 
\[
(S,\int H^{t}\frac{\delta _{l}}{\delta U^{t}}%
F_{m,n})=(n+1)F_{m-1,n+1}+mF_{m,n}\hspace{0.01in}+\int (-1)^{\varepsilon
_{s}\varepsilon _{t}}H^{t}U^{s}\frac{\delta _{l}^{2}}{\delta U^{t}\delta
H^{s}}(F_{m-1,n+1}-F_{m,n}). 
\]
Dividing by $n+m$ and summing over $m$ and $n$, we obtain 
\[
(S,\sum_{n+m>0}\frac{1}{n+m}\int H^{t}\frac{\delta _{l}}{\delta U^{t}}%
F_{m,n})=\sum_{n+m>0}F_{m,n}\hspace{0.01in}, 
\]
therefore (\ref{tesi}) holds with 
\begin{equation}
\chi ^{\prime }=\int H^{t}\frac{\delta _{l}}{\delta U^{t}}\sum_{n+m>0}\frac{%
F_{m,n}}{n+m}.  \label{singu}
\end{equation}
Clearly, if $\chi $ is local, so is $\chi ^{\prime }$. This concludes the
proof.

\bigskip

We can extend part of this theorem to gauge-semitrivial subsystems. First
observe that the argument given in the proof separately applies to each
``level'' $m+n=\ell =$constant, because no operation we have made mixes
different $\ell $s. This fact is due to the assumption that the system is
gauge-trivial. In a gauge-semitrivial subsystem, instead, the expressions $%
(S,\Phi _{r})$ are allowed to depend on $H$ and $U$, therefore $F^{\prime }$
raises the level. Attention must be paid to corrections coming from higher
and lower levels when attempting to extend the proof given above.

Call $\chi _{\ell }(\Phi)$ a functional that does not contain levels smaller
than $\ell >0$ and is such that $(S,\chi _{\ell })=0$. Let us repeat the
proof of the previous theorem focusing on level $\ell $. Since no
corrections can come from lower levels, the derivation works up to
corrections of levels $\ell +1$ or higher. We obtain the result that there
exist local functionals $\tilde{\chi}_{\ell }^{\prime }$ and $\chi _{\ell +1}
$ such that $\chi _{\ell }=(S,\tilde{\chi}_{\ell }^{\prime })+\chi _{\ell +1}
$. Clearly, $(S,\chi _{\ell +1})=0$, so we can repeat the argument for $\chi
_{\ell +1}$. Proceeding inductively, we conclude that there exists a local
functional $\chi _{\ell }^{\prime }$ such that $\chi _{\ell }=(S,\chi _{\ell
}^{\prime }) $. This derivation works for every $\ell >0$, but cannot be
extended to $\ell =0$, because formula (\ref{singu}) is singular in that
case. Therefore,

\begin{theorem}
If $H$-$U$ is a gauge-semitrivial $\Phi $-subsystem and $\chi (\Phi )$ is a
local functional of the fields that vanishes at $H=U=0$, such that $(S,\chi
)=0$, then there exists a local functional $\chi ^{\prime }(\Phi )$ such
that 
\[
\chi (\Phi )=(S,\chi ^{\prime }). 
\]
\label{theo8}
\end{theorem}

When we want to generalize the arguments of this appendix to functionals $%
\chi$ of both fields and sources, we must pay attention to a caveat. Both
theorems generalize if we can include the sources inside $\Phi _{r}$,
because the proofs never use the fact that $\Phi _{r}$ contains only fields.
The notion of gauge-semitriviality remains unchanged, so theorem \ref{theo8}
generalizes straightforwardly. Instead, the notion of gauge-triviality does
change, since the requirement that $(S,\Phi _{r})$ are independent of $H$
and $U$ would imply the requirement that $(S,K_{A})$ satisfy the same
property. In general, this is not even true for the subsystem $\bar{C}$-$B$
and the action (\ref{actr}), because $(S,K_{A})$ can depend on $\bar{C}$ and 
$B$, which makes the subsystem $\bar{C}$-$B$ gauge-semitrivial. At any rate,
for the needs of this paper it is sufficient to generalize theorem \ref
{theo8}:

\begin{theorem}
If $H$-$U$ is a gauge-semitrivial $\Phi $-subsystem and $\chi (\Phi ,K)$ is
a local functional that vanishes at $H=U=0$, such that $(S,\chi )=0$, then
there exists a local functional $\chi ^{\prime }(\Phi ,K)$ such that 
\begin{equation}
\chi (\Phi ,K)=(S,\chi ^{\prime }).  \label{tesi9}
\end{equation}
\label{theo9}
\end{theorem}

\end{document}